\documentclass[aps,prl,twocolumn,showpacs,superscriptaddress]{revtex4-1}  
\usepackage{graphicx}  
\usepackage{color}
\usepackage{amsmath}
\usepackage{gensymb}
\usepackage{ulem}
\usepackage{float}
\usepackage[utf8]{inputenc}
\usepackage{placeins}

\begin{document}

\title{DNA versus RNA - which shows higher electronic conduction?}

\author{Abhishek Aggarwal}
\affiliation{Center for Condensed Matter Theory, Department of Physics, Indian Institute of Science, Bangalore-560012, India}

\author{Saientan Bag}
\affiliation{Center for Condensed Matter Theory, Department of Physics, Indian Institute of Science, Bangalore-560012, India}

\author{Ravindra Venkatramani}
\affiliation{Department of Chemical Sciences, Tata Institute of Fundamental Research, Colaba, Mumbai-400005, India}

\author{Manish Jain}
\affiliation{Center for Condensed Matter Theory, Department of Physics, Indian Institute of Science, Bangalore-560012, India}

\author{Prabal K. Maiti}
\email{maiti@iisc.ac.in}
\affiliation{Center for Condensed Matter Theory, Department of Physics, Indian Institute of Science, Bangalore-560012, India}
\date{\today}

\begin{abstract}
In this study, we compare the charge transport properties of multiple (double stranded) dsRNA sequences with corresponding dsDNA sequences.
	Recent studies have presented a contradictory picture of relative charge transport efficiencies in A-form DNA:RNA hybrids and dsDNA.
	Using a multiscale modelling framework, we compute conductance of dsDNA and dsRNA using Landauer formalism in coherent limit and Marcus-Hush theory in the incoherent limit.
 We find that dsDNA conducts better than dsRNA in both the charge transport regimes.
 Our analysis shows that the structural differences in the twist angle and slide of dsDNA and dsRNA are the main reasons behind the higher conductance of dsDNA in the incoherent hopping regime.
 In the coherent limit however, for the same base pair length, the conductance of dsRNA is higher than that of dsDNA for the morphologies where dsRNA has smaller end-to-end length relative to that of dsDNA.
\end{abstract}
	 
\maketitle

\section{Introduction}
In recent years, the topic of DNA electronic conductance has gained much attention.
 DNA mediated electronic charge transport has been found to have biological implications\cite{tse2019effective}, relevant for processes such as redox switching of [4Fe4S] clusters found in DNA processing enzymes\cite{boon2003dna,o20174fe4s,bartels2017electrochemistry} and DNA damage \cite{merino2008biological,boal2005electrochemical}.
 The study of charge transport in nucleic acids is also relevant for assaying genetic materials\cite{li2018detection,aggarwal2018remarkable}.
 Self-assembly, multiple mechanisms of charge transport\cite{genereux2009mechanisms} and sequence-based tenability are some of the properties of dsDNA which make it a promising candidate for molecular electronics apart from its biological significance\cite{wang2018dna}.
 Several experimental\cite{xiang2017gate,li2016comparing,li2016long,kratochvilova2010charge, artes2015conformational,li2018detection} works have studied the charge transport properties of dsDNA.
 A variety of theoretical approaches have been developed to study charge transport properties of DNA\cite{ wolter2017microsecond, beratan1998electron, beratan2009steering, voityuk2007fluctuation,nitzan2003electron, jortner1998charge}.
 Previous studies have invoked incoherent hopping transport mechanisms\cite{bag2016dramatic,berlin2001charge} and coherent transmission of charges\cite{guo2016molecular,li2018detection} to explain the charge transport properties of DNA of various lengths.
 In contrast, the dsRNA have not yet received much attention and only a few studies exist\cite{wong2017hole,wu2016carrier} which provide some understanding of the dsRNA electronic charge transport properties.
 RNA is a macro-molecule comprising repeated stacks of nucleobases formed by either AU (UA) or GC (CG) pairs coupled via hydrogen bonds\cite{cuniberti2007tight}.
 In addition of having Uracil instead of Thymine, RNA also has different backbone than DNA.
 While DNA has a deoxyribose and phosphate backbone, RNA has a ribose and phosphate backbone.
 The conformation of a nucleic acid depends on the sequence, the environment of the NA chain\cite{kypr2009circular} and whether it is RNA or DNA\cite{gyi1996comparison}.
 Like dsDNA, RNA also exists in double stranded form, although dsDNA exists in B-form whereas dsRNA exists in A-form.
 Due to these similarities and dissimilarities of RNA with DNA, some obvious questions arise.
 Can dsRNA be used in molecular electronics? If yes, which one will conduct better- DNA or RNA?

 Recently, Tao et al.\cite{wu2016carrier} reported similar charge mobilities in both DNA and RNA and found that the hole mobilities are higher than the electron mobilities.
Yuanhui et al.\cite{li2016comparing,li2016long} compared the charge-transport properties of guanine-rich RNA/DNA hybrids to dsDNA duplexes with identical sequences experimentally and reported higher conductance and decay constants for A-form RNA/DNA hybrids relative to B-form dsDNA.
Wong et al.\cite{wong2017hole} reported hole transport properties of  A-form RNA/DNA hybrid duplexes to compare with B-form dsDNA duplexes.
They found a shallow distance dependence of hole transport rates in both types of duplexes but a lower yield for hole transport in DNA:RNA hybrids relative to dsDNA duplexes.
These contrasting results in the literature invoke a need for a more detailed examination  of the charge transport phenomenon in nucleic acids.
Moreover, the electronic charge transport properties of dsRNA cannot be estimated from RNA/DNA hybrids or A-form dsDNA as their conformational properties are different\cite{salazar1993dna}.
Through this study, we attempt to fill this gap.

The aim of the present study is to characterize the charge transport properties of A-form dsRNA and compare them to B-form dsDNA.
The electronic charge transport in nucleic acids has been previously described using three mechanisms: coherent tunneling\cite{qi2013unified}, intermediate tunneling-hopping\cite{renaud2013between} and incoherent hopping\cite{bag2016dramatic}.
In this work, we examine two different charge transport mechanisms: a) thermally induced hopping charge transport mechanism described by Marcus-Hush formalism\cite{marcus1993electron,hutchison2005hopping} and b) coherent transport mechanism described by Landauer formalism\cite{qi2013unified}.
Both the mechanisms involve completely different physics and treat the charge transport properties in two extreme limits: diffusive and coherent.
Previous studies present evidence for both the incoherent and coherent mechanisms in nucleic acids\cite{bag2016dramatic,qi2013unified,genereux2009mechanisms}.
Studies have further showed the transition from hopping regime to tunneling regime in the charge transport in DNA by varying the molecular length and sequence\cite{li2016thermoelectric}.
The exact mechanism of charge transport dominant in nucleic acids has attracted considerable debate\cite{bag2016dramatic,qi2013unified,genereux2009mechanisms,xiang2015intermediate,li2016comparing,li2016long} in the literature.
In this work, we do not attempt to resolve this issue, instead we present the charge transport properties using both the mechanisms and justify the results using different structural parameters.

We compute the charge transport properties using a multiscale modelling framework which combines molecular dynamics simulations and first principle calculations for determining the structural and electronic properties of nucleic acids.
 Additionally, the Kinetic Monte Carlo simulation is used in the hopping mechanism.
 The effect of sequence and base pair (bp) length is also studied here.
 We also compute the structural parameters such as rise, twist and inclination angle for both DNA and RNA sequences which are expected to influence the charge transport in the hopping case.

 We find that the dsDNA conducts better than dsRNA independent of the charge transfer mechanism.
 The difference in conductance of dsRNA and dsDNA in the incoherent hopping regime can be attributed to the different helical structural parameters of the two nucleic acids.
 While helical dsRNA is on average more compact relative to dsDNA of the same base pair length, the coherent conductance of dsDNA is higher than that of dsRNA of the same end-to-end length.
 However, the smaller end-to-end length of dsRNA relative to dsDNA leads to the experimentally observed higher conductance of the former.

The paper is organized as follows: In the Methodology section, we present the multiscale modelling framework used for the current calculation.
Detailed description of MD simulations and the formulations describing incoherent hopping and coherent transport are discussed in this section.
In the Results section, we compare the results obtained for dsDNA and dsRNA for both charge transport mechanisms and correlate the results to various structural parameters.
In the Conclusions section, we discuss the implications of our results and connections to experiments.

\begin{figure}[h!]
        \includegraphics[width=\columnwidth]{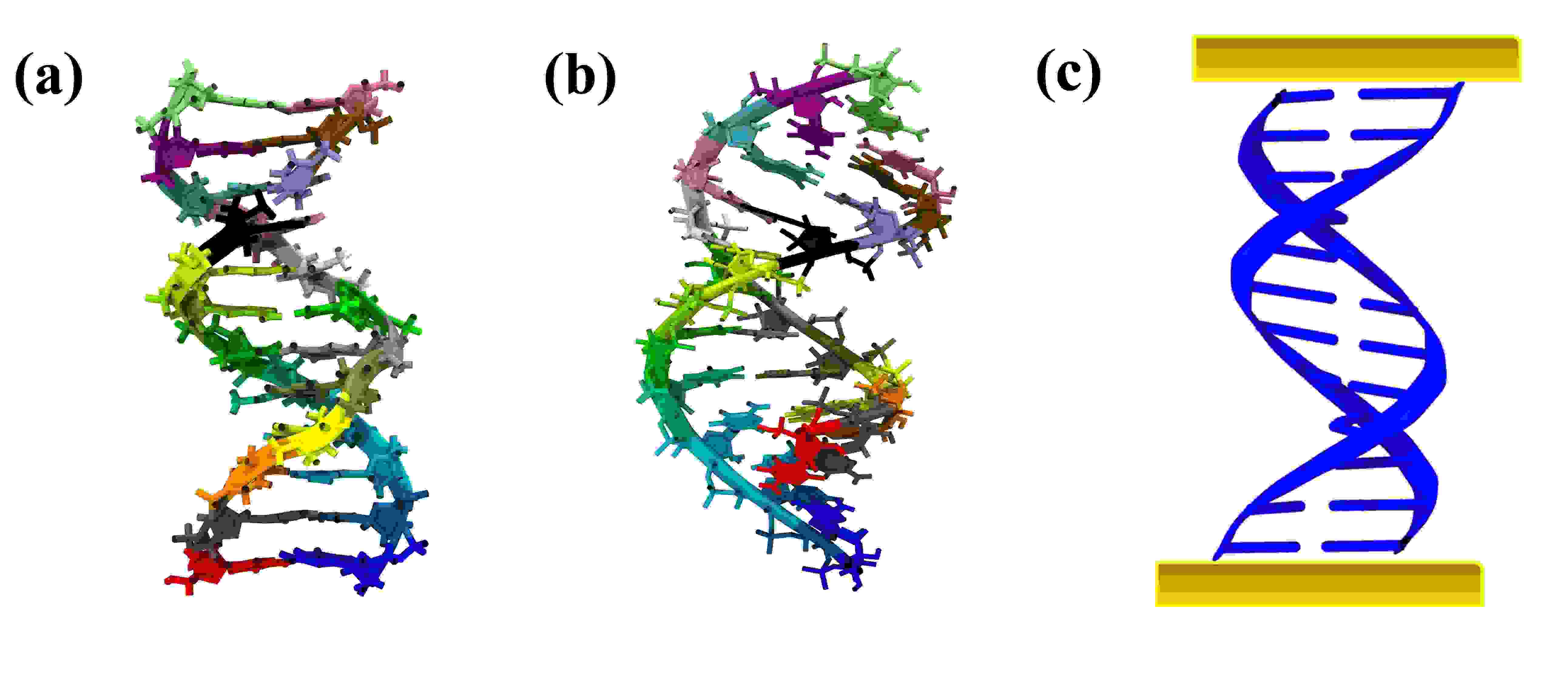}
        \caption{Initial structure of (a) B-form DNA sequence d-(CGCGAATTCGCG). (b) A-form dsRNA d-(CGCGAAUUCGCG) used for MD simulation. (c) Schematic diagram showing DNA with electrodes. DNA/RNA are shown in blue, while the electrodes are shown in yellow colour.}
    \label{fig:fig_1}
\end{figure}

\section{Methodology}
\subsection{MD Simulation} 
As a first step of the multiscale modeling framework, we performed all-atom molecular dynamics (MD) simulation of the nucleic acids to predict a realistic solvated structural ensemble of the nucleic acids.
 The initial structures of the duplex RNA in A-form and duplex DNA in B-form for MD simulations were generated using the NAB\cite{duke2016amber} module available in the AMBER\cite{duke2016amber} suite of programs.
 We then simulated dsRNA with lengths of 12, 14, 16, 18 and 20 base pairs, using the sequence $d-(CGCGA_{n}U_{n}CGCG)$ with n = 2 to 6 and $d-(CG)_{n}$ (with n = 2 to 6) dsDNA and dsRNA sequences to examine the length dependence of charge transport properties of nucleic acids.
 In addition, to study the effect of nucleic acid sequence on our results comparing dsDNA and dsRNA charge transport in incoherent and coherent regimes, we also simulated four other 12 bp long dsRNA and the corresponding dsDNA sequences.
 These include d-(CGCGCGCGCGCG), d-(CCCCCCCCCCCC), d-(AAAAAAAAAAAA) and d-(AUAUAUAUAUAU) (d-(ATATATATATAT) for dsDNA).

 Each dsDNA or dsRNA structure was solvated in a box of water using TIP3P water model\cite{jorgensen1983comparison} using xleap module of AMBER 16\cite{duke2016amber}.
 The water box dimensions were chosen to ensure 15 \AA $ $ solvation of the nucleic acid in each direction.
 Appropriate number of Na+ counter ions were added to neutralize the negative charge of phosphate backbone of the nucleic acid chain.
  We used  a combination of ff99bsc0 and OL3\cite{perez2007refinement,zgarbova2011refinement} for RNA, whereas for B-form DNA, we used  ff99bsc0 and OL15\cite{zgarbova2013toward,zgarbova2015refinement} force fields.
 The system was equilibrated at a temperature of 300 K and a pressure of 1 bar.
 We performed a production run of 100 ns for all the  sequences using the PMEMD code available in AMBER\cite{duke2016amber} package.
 For more details about the simulation protocol, please refer to our previous work\cite{aggarwal2018remarkable}.

Structural snapshots along the trajectories were saved after every 2 ps.
For each sequence, we used 50 equilibrated snapshots for charge transport calculations in the hopping charge transport mechanism whereas for the coherent mechanism, 25 snapshots were used for the calculations.
The initial NAB\cite{duke2016amber} generated structures of B-DNA and A-RNA are shown in fig. \ref{fig:fig_1}(a) and fig. \ref{fig:fig_1}(b).
 Fig. \ref{fig:fig_1}(c) shows a schematic diagram of a nucleic acid chain with electrodes connected to its terminals.
 The structural analysis of the dsDNA/dsRNA is done using cpptraj module\cite{duke2016amber}.

\subsection{The Hopping Transport}
We use the Semi-Classical Marcus-Hush\cite{deng2015quantitative,marcus1993electron} formalism to calculate the current through the double stranded nucleic acids.
 In this theory, charge transport is described as incoherent hopping of charge carriers between charge hopping sites.
 Several theoretical and experimental investigations have demonstrated that the charge transport in nucleic acids is mediated by stacked nucleobases through strong $\pi-\pi$ interactions\cite{endres2004colloquium,aggarwal2018remarkable}.
 In this mechanism, we remove the backbone atoms from both the dsDNA and dsRNA and use only the nucleobases capped with hydrogen  to satisfy valencies of the dangling bonds for further calculations and optimizations.
 In Marcus-Hush formalism, the charge transfer rate $\omega_{ik}$ from $i_{th}$ charge hopping site to the $k_{th}$ hopping site is given by\cite{marcus1993electron,aggarwal2018remarkable}
\begin{equation}
\omega_{ik} = \frac {2\pi} {\hbar} \frac {|J_{ik}^{2}|}{\sqrt{{4 \pi}{\lambda_{ik} k_{B} T}}} exp [\frac{-{(\Delta G_{ik}-\lambda_{ik}})^2}{4\lambda_{ik} k_B T}]
\label{eq:hopping_rate}
\end{equation}

 where $J_{ij}$ is the transfer integral\cite{baumeier2010density}, defined as

\begin{equation}
J_{ik} = <\phi^{i}|H|\phi^{k}>
\end{equation}

 Here, $\phi^{i}$ and $\phi^{k}$ are the diabatic wave functions localized on the $i^{th}$ and $k^{th}$ sites respectively.
 Both the hole and electron current are calculated in this study.
 In the case of hole transport, the Highest Occupied Molecular Orbital (HOMO) are used as the diabatic wave function whereas, for electron transport, Lowest Unoccupied Molecular Orbital (LUMO) are used.
 To account for the effect of dynamic disorder arising due to the thermal fluctuations, electronic couplings between all possible nearest neighbors charge hopping base pairs (fig. S8 of ESI) are computed for 50 nucleic acid snapshots sampled from MD simulations. 
 H is the Hamiltonian for the two-site system between which the charge transfer takes place.
 $\lambda_{ik}$ is the reorganization energy.
 $\Delta G_{ik}$ is the free energy difference between two sites, $h$ is the Planck's constant, $k_B$ is the Boltzmann constant, and $T$ is the absolute temperature.

 The reorganization energy, $\lambda_{ik}$, has two parts : inner sphere reorganization energy and outer sphere reorganization energy.
 Inner sphere reorganization energy takes into account the change in nuclear degrees of freedom when the charge transfer takes place between one charge hopping site to another.
 This is defined as\cite{rhle2011microscopic, aggarwal2018remarkable,bag2016dramatic},

\begin{equation}
\lambda_{ik}^{int} = U_{i}^{nC} - U_{i}^{nN} + U_{k}^{cN} - U_{k}^{cC}
\end{equation}
 $U_{i}^{nC} (U_{i}^{cN})$ is the internal energy of neutral (charged) base in charged (neutral) state geometry.
 $U_{i}^{nN} (U_{i}^{cC})$ is the internal energy of neutral (charged) base in neutral (charged) state geometry.
 Whereas, the reorganization of the environment as the charge transfer occurs is considered using the outer sphere reorganization.
 This parameter should have a similar value for dsDNA and dsRNA as the solvation conditions for the two systems were the same for all calculations.
 For simplicity, we set the outer-sphere reorganization to 0 eV, in our calculations.
 Using non-zero values of the outersphere reorganization energy (up to 1 eV) does not change the relative conductance trend of dsDNA and dsRNA (fig. S9 of ESI).

Free energy difference between two charge hopping sites consits of two parts- internal free energy difference and external free energy difference. 
 The internal free energy difference is taken as the difference between internal energies of the two hopping sites as\cite{rhle2011microscopic, aggarwal2018remarkable,bag2016dramatic}
\begin{equation}
	\Delta G_{ik}^{int} =  \Delta U_{i} -  \Delta U_{k}
\end{equation}

\begin{equation}
        => \Delta G_{ik}^{int} = (U_{i}^{cC} - U_{i}^{nN}) - (U_{k}^{cC} - U_{k}^{nN})
\end{equation}

 , where $\Delta U_{i(k)}$ refers to the adiabatic ionization potential (or eectron affinity) of base $i(k)$ and $\Delta U_{i(k)}^{cC(nN)}$ is the total energy of the base $i(k)$ in charge (neutral) state and geometry. 
 The ionization energies and electron affinities of dsDNA and dsRNA nucleobases are provided in table 2 of ESI.
 The external free energy difference is taken as the potential difference between the two hopping sites as described in ESI.

 The calculation of transfer integrals and reorganization energies are performed using density functional theory (DFT) which have been carried out with M062X/6-31g(d) functional level of theory using Gaussian09\cite{frisch2009gaussian} software package.
Polarizable Continuum Model (PCM)\cite{tomasi2005quantum} is used in the calculations to consider the effect of solvation arising due to surrounding water medium of the base pairs.
 VOTCA-CTP\cite{rhle2011microscopic,baumeier2010density} software package is also used to calculate the transfer integral values for all possible base pairs.

 \subsection{Kinetic Monte Carlo}
 Once the charge hopping rates are obtained for all possible base pairs, Kinetic Monte Carlo\cite{chatterjee2007overview,rhle2011microscopic} (KMC) method is used for the calculation of V-I characteristics.
 In KMC algorithm, the time evolution of the system is described by solving the master equation of probabilities.\cite{aggarwal2018remarkable,bag2016dramatic,rhle2011microscopic}.
In this scheme, a random charge hopping site, $i$, is assigned a unit charge at the initial time $t = 0$.
 To calculate the  waiting time $\tau$ at charge hopping site $i$, we use the relation\cite{aggarwal2018remarkable,bag2016dramatic,rhle2011microscopic}:

\begin{equation}
\tau = -\omega_i^{-1} \ln (r_1) 
\end{equation}

where $\omega_i = \sum_{j=1}^n \omega_{ij}$ is the sum of the charge hopping rates for all the possible hopping sites, $j$ from site $i$, $n$ is the total number of charge hopping sites available for charge at site $i$ and $r_1$ is a uniform random number between 0 and 1.
 After the calculation of the waiting time, the total time is then updated as $t = t + \tau$.
 The hopping site, the $j$ for which  $\frac{\sum_j \omega_{ij}}{\omega_i} $ is largest and $\leq r_2$, is chosen as the site where the charge hops next.
Here, $r_2$ is another uniform random number between 0 and 1.
The above condition ensures that the site $j$ is chosen with probability $\frac {\omega_{ij}}{\omega}$.
 After this, we update the position of the charge and repeat the above process which provides the probabilities for each site.
 The current is then computed by solving the following master equation of probabilities\cite{livshits2014long},

\begin{equation}
I_{bp } = -e [\sum_i (P_{b_1} \omega_{b_1i} - P_{i} \omega_{ib_1}) + \sum_i (P_{b_2} \omega_{b_2i} - P_{i} \omega_{ib_2})]
\end{equation}

 Here, $e$ is the unit electric charge, $i$ stands for all the possible hopping sites which are in the direction of flow of current, $b_1$ and $b_2$ are the base stacks of base pair $bp$.
 Hence, the mean current is average over all base pairs, $I = < I_{bp} >$.

\subsection{The Coherent Transport}

To calculate the charge transport properties using coherent tunneling mechanism, we use a methodology which consists of the following three steps: a) ab initio calculations to obtain the Hamiltonian matrix for the full dsDNA/dsRNA system, b) use of non-equilibrium Green's function (NEGF) method to get the transmission through the molecule, and finally c) obtaining the V-I characteristics using Landauer formalism.
This framework has been used in several previous works to study the charge transport properties of various organic systems.\cite{seth2017conductance, venkatramani2010evidence, xing2010optimizing}

We consider 25 nucleic acid structures sampled from the last 2 ns of 100 ns long MD simulated trajectory in our calculations. 
 To obtain the Hamiltonian matrices for these structures, we use GAUSSIAN 09\cite{frisch2009gaussian} adopting the semi-empirical method PM3\cite{stewart1989optimization}.
The semi-empirical method is used to reduce the computational cost of the calculations.
 The Fock matrix obtained after semi-empirical calculation, which is in the basis of atomic orbitals, is taken as the Hamiltonian matrix for subsequent calculations.

 To obtain the transmission through the molecule, the NEGF framework is used, in which the molecular Green's function is modified due to atomic contacts with virtual electrodes.
The Green's function is given by\cite{seth2017conductance,nitzan2003electron}:
\begin{equation}
        G(E) = \frac{1}{(E\textbf{I}-\textbf{H}-\Sigma^{l}-\Sigma^{r})}
\end{equation}

where, $H$ is the Hamiltonian of the isolated molecule.
The self-energies $\Sigma^{l}$ and $\Sigma^{r}$ describe the broadening and shifting effects of the left (l) and right (r) electrodes respectively, on the molecular energies.
 In our calculations, only imaginary part of the self-energy matrix is considered.
 The transmission probability for charge transport from one electrode to the other electrode over all pathways is given by:
\begin{equation}
        T(E) = \Gamma^{l}G\Gamma^{r}G^{\dagger}
\end{equation}

Here, $\Gamma^{l}$ and $\Gamma^{r}$ are the broadening matrices given by $\Gamma = i[\Sigma - \Sigma^{\dagger}]$.
 Several studies on molecular conductance\cite{qi2013unified,seth2017conductance,gutierrez2006inelastic} have been carried out under these approximations wherein the electrode atoms are not explicitly modelled, instead the effect of the electrodes has been considered using the broadening matrices.
 We assume that the electrodes electronically couple only to the terminal base pairs and add the broadening parameter on all the atomic orbitals representing the terminal base pairs`  atoms only except for the backbone atoms.
 Hence, the elements of the broadening matrices are given as $\Gamma_{ij}  = 0.1$ eV, for the terminal base pair atomic orbitals and $i = j$, and is taken as 0 eV otherwise.

The choice for the broadening matrices was fitted to reproduce the magnitude of experimental currents.
Using the above formalism and parameters, we get the value of transmission coefficient for a range of energy values.
 After that, the Landauer expression \cite{datta1997electronic,landauer1987electrical,xin2019concepts} is used to get the value of current $I$ at a given voltage $V$:
\begin{equation}
        I = \frac{2e^{2}}{h}\int_{\infty}^{-\infty} dE[f(E + \frac{eV}{2}) - f( E - \frac{eV}{2})]T(E)
\end{equation}

Here, $f(E)$  is the Fermi energy function given by:
\begin{equation}
        f(E) = \frac{1}{1 + exp((E-\mu)/k_{B}T)}
\end{equation}

Here, $k_{B}$ is the Boltzmann constant and $T$ is the temperature taken as 300 K and $\mu$ is the chemical potential of the electrodes.
 We find that the Fermi level of the dsDNA system increases by 0.36 eV upon the attachment of gold electrodes in our other yet unpublished work.
 Hence, the Fermi level in our calculations has been taken to be 0.36 eV above the HOMO energy level of each DNA/RNA snapshot.
 Please note that these calculations may overestimate resonant transport as the decoherence effects are neglected here\cite{venkatramani2014breaking}.
The above formalism is applied to 25 randomly chosen snapshots from last 2 ns trajectory of 100 ns long MD simulation of both dsDNA as well as dsRNA and the averaged V-I characteristics are presented in the results section.

\section{Results and discussion}
\subsection{Hopping Mechanism}

\begin{figure}[h!]
        \centering
        \includegraphics[width=\columnwidth]{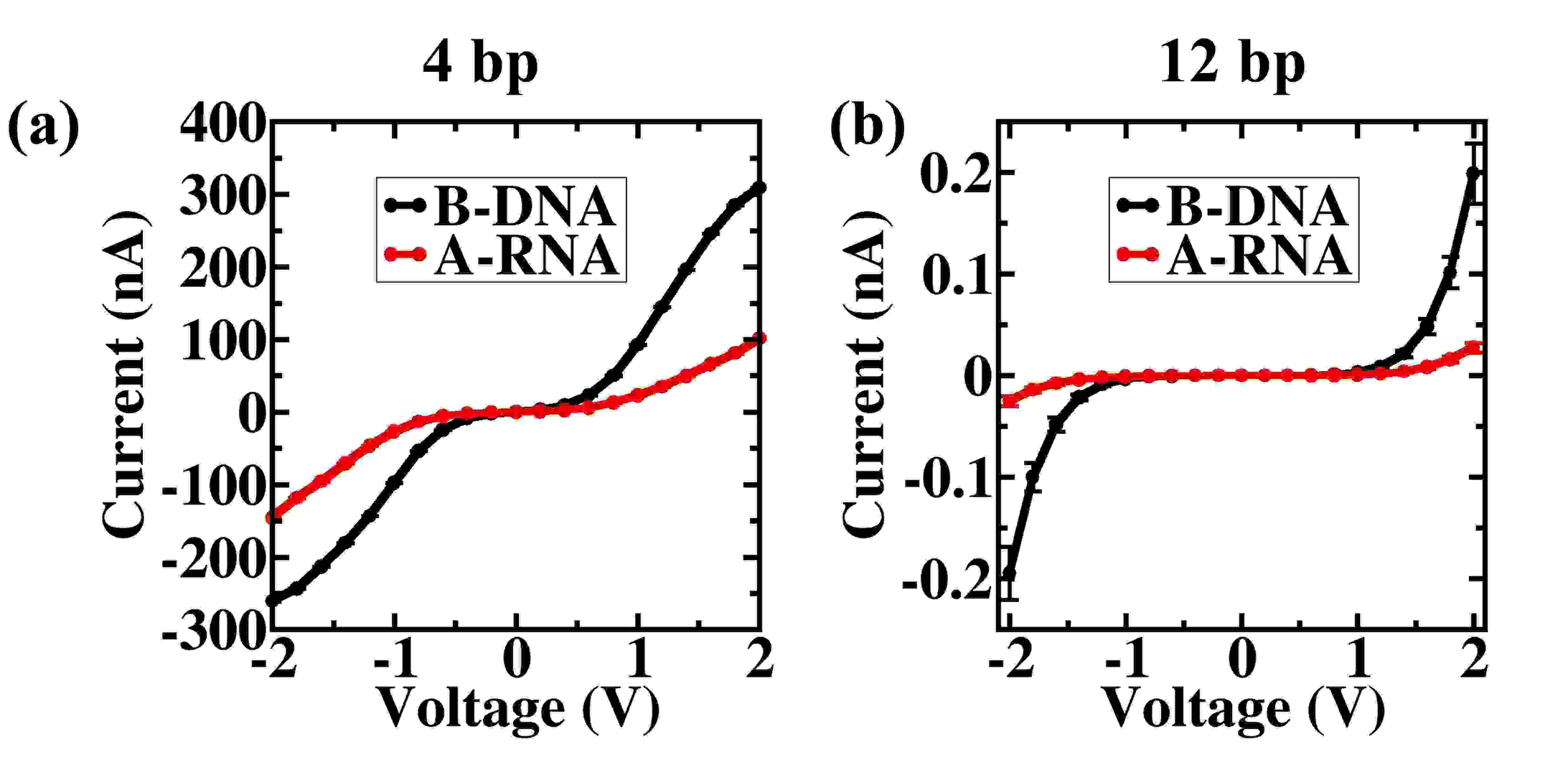}
        \caption{V-I characteristic curve for (a) 4bp d-(CGCG) B-DNA and A-RNA sequences and (b) 12 bp d-(CGCGAATTCGCG) B-DNA and A-RNA sequences using hopping transport mechanism. The hole current for 4 bp case is $\mu$A, whereas for 12 bp sequences, it is in nA range. In both the cases, dsDNA conducts better than dsRNA.}
        \label{fig:fig_2}
\end{figure}

The average V-I characteristic curve for 4 bp A-RNA d-(CGCG) and B-DNA d-(CGCG) are shown in fig. \ref{fig:fig_2}(a).
For comparison, we also show the V-I characteristic curve for the 12 bp dsDNA and dsRNA d-(CGCGAATTCGCG) in fig. \ref{fig:fig_2}(b).
Clearly, A-RNA has lower conductance than B-DNA, although the V-I characteristics are similar i.e. the current is negligible for very low voltages and grows non-linearly above a threshold voltage.
The current is of the order of $\mu$A for 4 bp sequences, while for 12 bp sequences, it is in nA range.
To understand the effect of length of RNA on the conductance in the hopping regime, we calculate the decay constant using the conductance vs base pair lengths of A-RNA $(d-(CGCGA_{n}U_{n}CGCG))$ as shown in fig. S7a of ESI.
The decay constant value is 0.16 \AA$^{-1}$ which is very small and indicates that the length dependence of conductance is weak.
A similar pattern has been found for dsDNA\cite{bag2016dramatic} in previous works as well.
The lower hole transport yield of A-form DNA:RNA hybrids than B-form dsDNA in the regime where the hole transport is weakly dependent on the length, is also seen in experiments before\cite{wong2017hole}.

To understand the effect of sequence on the conductance\cite{kratochvilova2010charge}, we obtained the V-I curve for various DNA and RNA sequences.
The results obtained are very similar, i.e. in each case B-form dsDNA conducts better than the corresponding A-form dsRNA (fig. S1 of ESI).
The V-I characteristics shown here are for the hole currents.
The electronic current is also calculated which is shown in fig. S2 of ESI.
The electronic current is 2 orders of magnitude smaller than the hole current, but for the electronic current as well, the trend remains the same, i.e. dsDNA conducts better than dsRNA.
The difference in the electronic and hole conductance is due to higher reorganization energies required for electrons relative to holes.
For example, the reorganization energy for hole transport between cytosine and guanine is 0.60 eV whereas for electron, it is 1.06 eV.
Readers are referred to ESI for more information on differences in reorganization energies for holes and electrons for different nucleobases.
So, higher hole current is not surprising and has been noted in several previous works\cite{wong2017hole,wu2016carrier, kratochvilova2010charge}.

As we have removed the backbone atoms from the calculations, the chemical properties of DNA and RNA are similar.
So, the difference in the conductance of DNA and RNA must be due to their structural differences.
So, in the following sections, we examined how the two nucleic acids species differ in terms of their structural parameters namely rise, helical rise, slide, inclination angle and twist.

\subsection{Structural Parameters}

\begin{figure}[h!]
    \centering
            \includegraphics[width=\columnwidth]{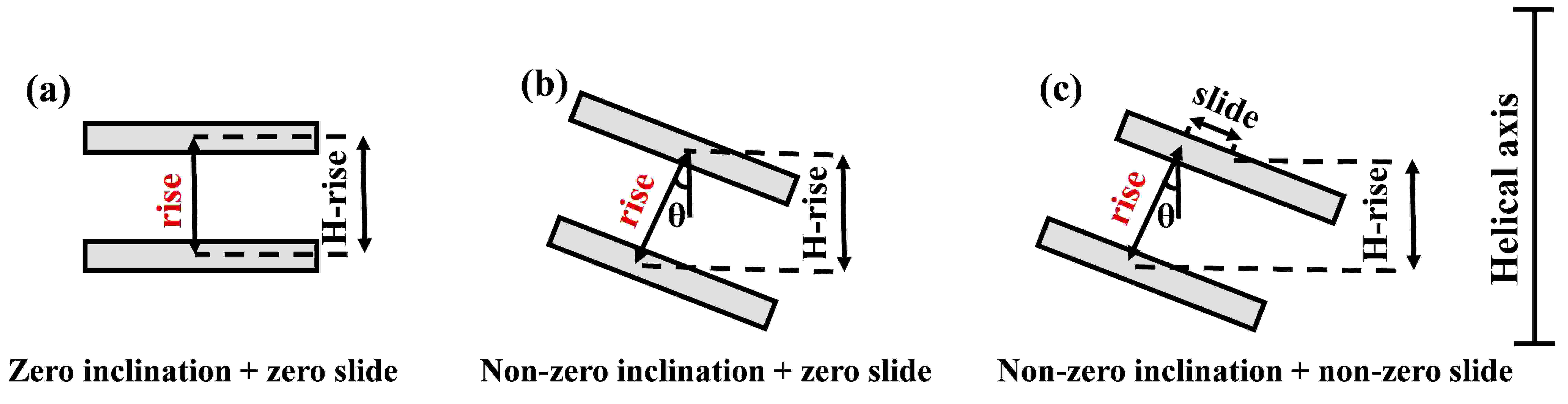}
        \caption{Schematic diagram describing the difference between the helical rise and rise of nucleic acids. (a) If the inclination angle and slide are zero, the bases are stacked one above other. (b) For a non-zero inclination angle, the bases are tilted with respect to the helical axis. (c) For a non-zero inclination angle as well as non-zero slide, one base is displaced with respect to other base in addition to the tilt. Thus, higher value of slide leads to larger distance between the two bases.}
    \label{fig:schematic_slide}
\end{figure}

To understand the structural differences between A-RNA and B-DNA, we plotted histograms of the helical parameters for both DNA and RNA.
 The inclination angle is a crucial parameter to understand the structure of a nucleic acid.
 This is defined as the angle between the helical axis and the rise vector\cite{bao2017understanding}. 
 As described in fig. \ref{fig:schematic_slide}, the helical rise is the distance between two consecutive base pairs along the helical axis, whereas rise is the inter base-pair translational parameter which measures the distance between the two bases in the base pair reference frame\cite{lu20033dna,blanchet2011curves}. 
 The difference in helical-rise and rise becomes larger as the inclination angle increases (fig. \ref{fig:schematic_slide}).
The relation between helical rise ($h-rise$) and rise can be understood in terms of inclination angle ($\theta$)  and slide as (fig. \ref{fig:schematic_slide}) \cite{bao2017understanding}:
\begin{equation}
h-rise = rise * cos\theta + slide * sin\theta
\end{equation}

\begin{figure}[H]
        \includegraphics[width=\columnwidth]{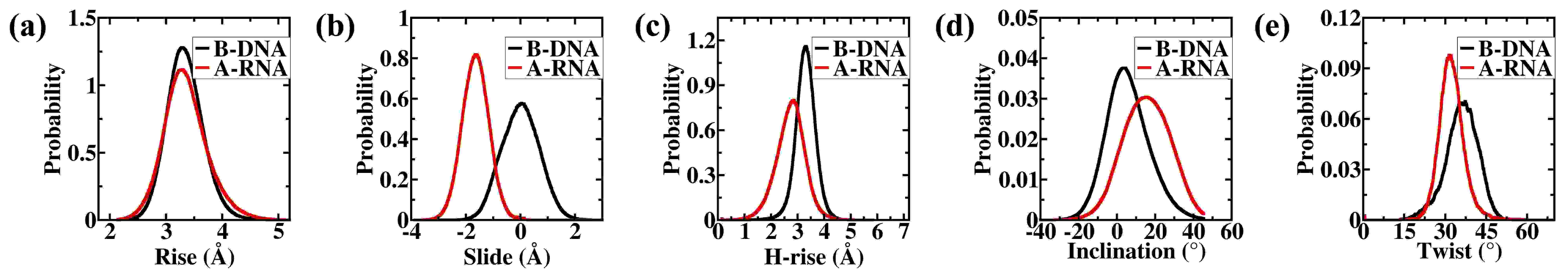}
      \caption{Distribution of helical parameters for d-(CGCGAATTCGCG) sequence for B-DNA and correspondingly d-(CGCGAAUUCGCG) for A-RNA. (a) Rise: Rise for A-RNA and B-DNA are almost same. (b) Slide: A-RNA has higher absolute value of slide than B-DNA. (c) Helical Rise: Although rise of B-DNA and A-RNA are similar, the helical rise of B-DNA is higher in magnitude than A-RNA. (d) Inclination angle: A-RNA has very high inclination angle than B-DNA. (e) Twist: A-RNA has lesser twist angle than B-DNA.}
    \label{fig:parameters}
\end{figure}

The comparison of structural parameters between A-RNA and B-DNA leads to a conclusion that although A-RNA has less helical rise than B-DNA, they have similar rise (fig. \ref{fig:parameters}).
Other observations are that the A-RNA has higher inclination angle and slide than B-DNA, whereas B-DNA has higher twist angle value than A-RNA.
The reason behind the difference in the helical rise of A-RNA and B-DNA is the larger inclination angle and larger slide in A-RNA than B-DNA due to which the helical rise becomes smaller in A-RNA (fig. \ref{fig:parameters}).
The charge transfer hopping rate is highly sensitive to the electronic coupling of the hopping sites.
So, any parameter that affects transfer integral will directly affect the conductance values.
We try to find the dependence of transfer integral values on various structural parameters of nucleic acids in the next section.

\onecolumngrid

\twocolumngrid

\subsection{Dependence of transfer integral values on nucleic acid structural parameters}
\begin{figure}[h!]
        \includegraphics[width=\columnwidth]{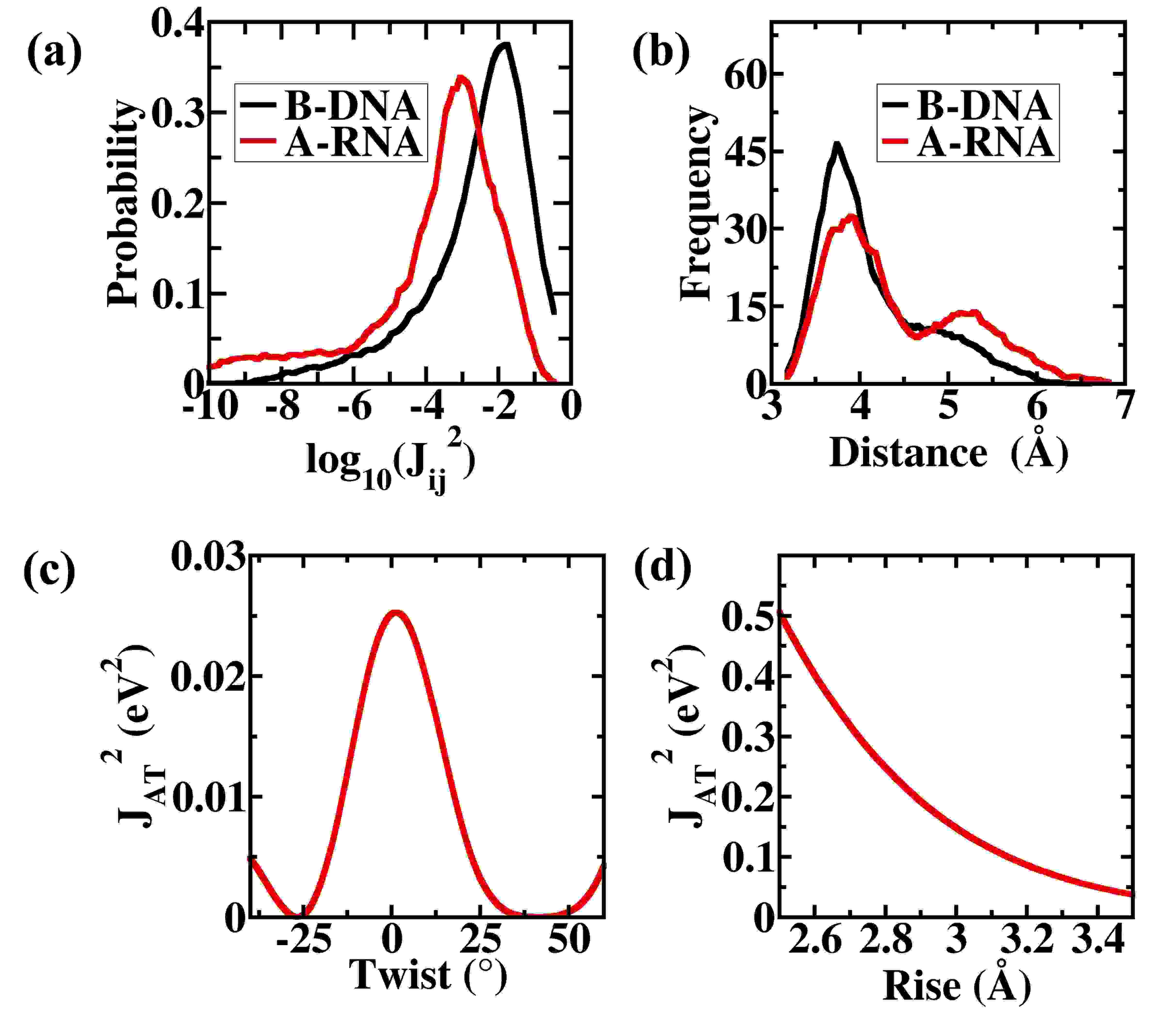}
	\caption{(a) Probability distribution of log of square of electronic coupling values of all possible hopping pairs of 50 morphologies of B-form dsDNA (d-(CGCGAATTCGCG)) and A-form dsRNA (d-(CGCGAAUUCGCG)).(b) Distribution graph of distance between consecutive base pairs of B-DNA and A-RNA showing there are more small-distanced bases in B-DNA than A-RNA, hence current is higher in B-DNA. (c) Graph showing variation of the square of transfer integral values with rise between two bases (A and T). Clearly, the transfer integral values decrease rapidly with increasing distances between bases. (d) Variation of square of transfer integral values with twist angle between A and T shows that the transfer integrals depend periodically on the twist angle values.}
    \label{fig:fig_5}
\end{figure}

The electronic coupling between two charge hopping sites is highly sensitive to their relative orientation as well as their relative distance\cite{voityuk2008conformations,voityuk2001charge}.
In case of dsDNA and dsRNA, computed structural parameters such as slide, inclination angle and twist angle impact the relative geometry of two bases in different ways.
For instance, the effect of higher slide in dsRNA is to increase the distance between the bases, whereas different twist angle is responsible for different relative orientations in dsDNA and dsRNA.
 The probability distribution of electronic couplings for all possible charge hopping sites of dsDNA and dsRNA shows that the electronic coupling values are higher for B-DNA than A-RNA (fig. \ref{fig:fig_5}(a)).
We present a distribution plot of center of mass (C.O.M.) distance between bases in A-form dsRNA of d-(CGCGAAUUCGCG) and B-form dsDNA d-(CGCGAATTCGCG) chain (fig. \ref{fig:fig_5}(b)) and find that the distance between bases in dsRNA is higher than B-form dsDNA.
The variation of transfer integral values with the distance between two bases A and T is presented in fig. \ref{fig:fig_5}(d)).
Clearly, the transfer integral decays exponentially with increase in rise.
This dependence plays a pivotal role in deciding the hopping rate between two charge localization sites and hence decides the conductance of a DNA or RNA chain\cite{venkatramani2011nucleic}.
This is reflected in the conductance values of these sequences.
We also plotted the variation of transfer integral values with twist angle (fig. \ref{fig:fig_5}(c)).
The transfer integral values vary periodically with twist angles.
Thus, the conductance of a nucleic acid chain can be estimated from its structural properties using charge hopping mechanism.
As a concluding remark, it can be said that the dsDNA conducts better than dsRNA in hopping regime due to their structural differences.
The dependence of transfer integral on slide and twist leads to a difference in the charge hopping rates between charge hopping sites leading to different charge transport properties of the two molecules.

\subsection{Effect of Disorder on the charge transfer properties}

\begin{figure}[h!]
        \includegraphics[width=\columnwidth]{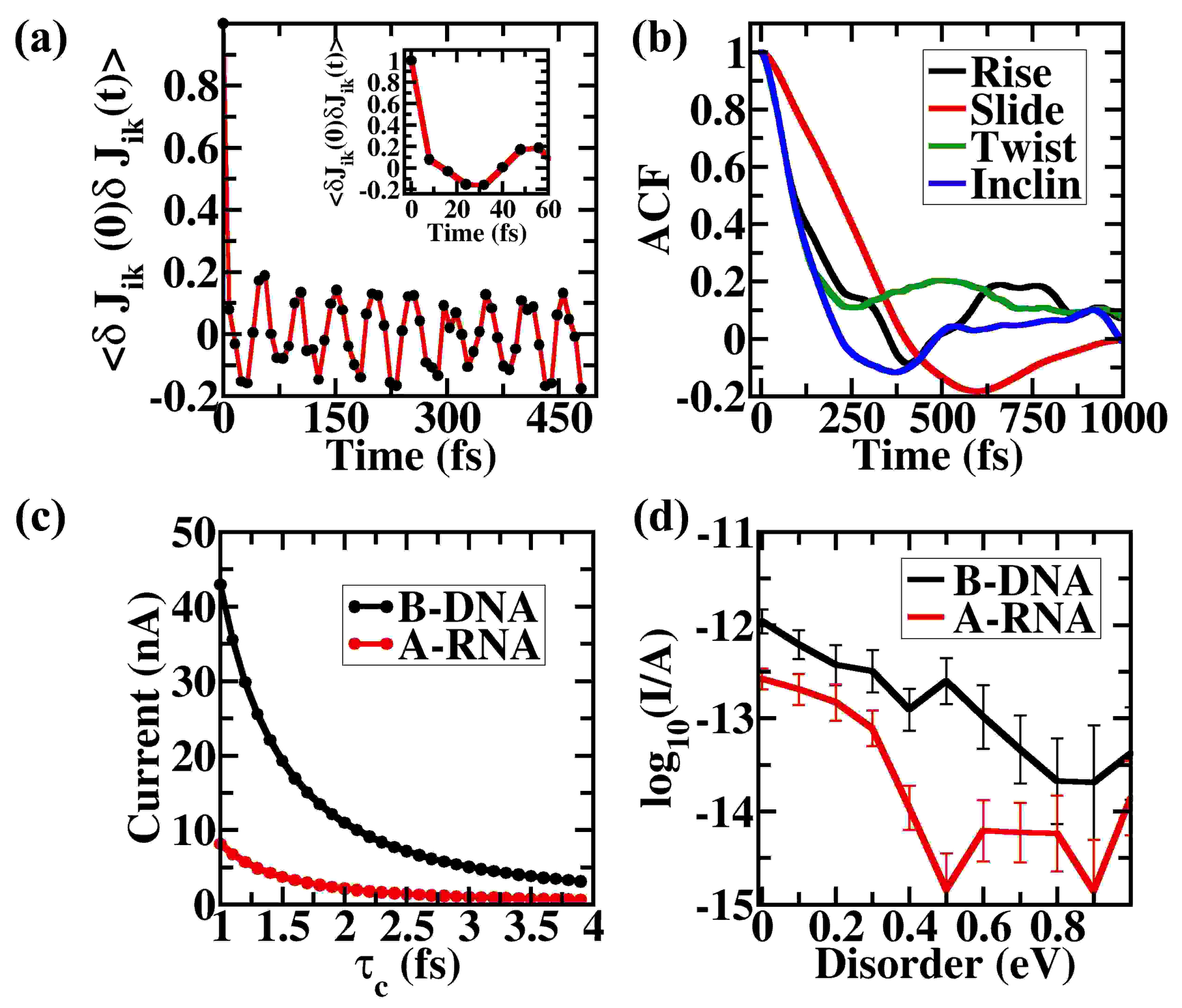}
	\caption{(a) Auto-correlation function for electronic couplings between the third G:C base pair of Drew-Dickerson dsRNA. Clearly, the electronic couplings decay rapidly on a ~10 fs timescale. The inset shows the zoomed auto-correlation function decay over the first 60 fs. (b) Auto-correlation function for various structural parameters for the same G:C base pair. Correlations for the structural fluctuations decay on sub-picosecond (100-500 fs) timescale. (c) The current at 2 V for B-form dsDNA and A-form dsRNA with disorder explicitly considered using the formalism by Troisi et al.\cite{troisi2003rate}. Clearly, dsDNA conducts better than dsRNA for any value of $\tau_{c}$. (d) Variation of the current at an applied potential bias of 1V with the amplitudes of the additional disorder incorporated to the site energy values.}
    \label{fig:fig_6}
\end{figure}

 The fluctuations in the geometry of the nucleic acid base pairs are known to be directly coupled to the charge transfer rate.\cite{troisi2003rate, troisi2002hole,berlin2008charge, jang2005theory,grozema2008effect, woiczikowski2009combined}.
 In our calculations, the structural fluctuations are automatically considered since we compute the electronic couplings for each possible base pair of 50 different morphologies taken from the MD simulated trajectories.
 This ensures that the effect of fluctuations of each structural parameter of the nucleic acid like rise, slide, twist and inclination angle is considered properly in the charge transport calculations.

In addition to the static disorder described above, dynamic fluctuations in nucleobase geometries (dynamic disorder) on the timescale of charge transport can impact hopping rates.
In order to account for the effect of dynamic disorder on the transport properties of dsDNA and dsRNA, we use the formalism described by Troisi et al.\cite{troisi2003rate}
Here, the correction in the hopping rate is given by the following term\cite{troisi2003rate}:

\begin{equation}
	\omega_{ik}^{(2)} = \omega_{ik}^{(0)} 2 \frac {\hbar^{2}} {\tau_{c}^{2}} \frac {(\lambda_{ik} + \Delta G_{ik})^2 - 2\lambda_{ik} k_B T}{(4\lambda_{ik} k_{B} T)^{2}} (1-\frac{<J_{ik}>^{2}}{<J_{ik}^{2}>})
\end{equation}

where $ \omega_{ik}^{(2)}$ is the correction term to the Marcus-Hush hopping rate expression as described in eq. \ref{eq:hopping_rate} and ${\tau}_{c}$ is the characteristic fluctuation time of electronic couplings.
The term $J_{ik}^{2}$ is replaced by $<J_{ik}^{2}>$ in the eq. \ref{eq:hopping_rate} as described in Ref. \cite{troisi2003rate}.
We calculate the current for a potential bias of 2 V for a range of ${\tau}_{c}$ values using the modified hopping rate values for each base pair and find that the relative trend of conductance of dsDNA and dsRNA does not change (fig. \ref{fig:fig_6}(c)).
This implies that the dsDNA conducts better than dsRNA regardless of the method of consideration of the dynamic disorder in the calculation.

To explicitly examine the effect of disorder in the site energies and electronic couplings on the relative conduction properties of dsDNA and dsRNA, we implement the formalism used by Siebbeles et al.\cite{kocherzhenko2009mechanism}.
 We introduce fluctuations in the site energies as uncorrelated stochastic processes in the KMC simulations.
 We sampled the additional fluctuations from a uniform distribution with amplitude, $A$, centered at zero as, $\delta E \in [-A,A] $ where $A$ varies between 0 and 1 eV.
Fig. \ref{fig:fig_6}d  shows the variation of the current at an applied potential bias of 1 V across dsDNA and dsRNA with the amplitudes of the additional disorder added to the site energy values. 
Notice that the conductance varies over several orders of magnitude with changes in the disorder amplitudes.
This is expected since the charge hopping rates depend exponentially on the site energy differences (eq. \ref{eq:hopping_rate}). 
 Despite this variability, the conductance of dsDNA is higher than that of dsRNA for all values of disorders in site energies considered here.  
 Similarly, introducing additional fluctuations to the electronic couplings of two hopping sites does not alter the relative electronic conduction properties of dsDNA and dsRNA (fig. S5).  
 Thus, our conclusion that B-form dsDNA shows better electronic conduction than A-form dsRNA in the incoherent hopping regime is unaltered upon addition of dynamic disorder in KMC parameters.

To understand the time scales involved in charge fluctuation as well as structural fluctuations, we calculated electronic couplings for the third C:G base pair for each snapshot taken after every 8 fs from a short simulation of 10 ps of Drew Dickerson dodecamer dsRNA.
Fig. \ref{fig:fig_6}(a) and fig. \ref{fig:fig_6}(b) shows the autocorrelation function of the electronic couplings and various structural parameters respectively.
Clearly, the timescale over which electronic coupling correlations decay is much smaller (<100 fs) than the timescale for fluctuations in nucleobase geometries (~picoseconds).
Thus, the effect of dynamics disorder appears to be minimal in these systems.
To summarize, dsDNA conducts better than dsRNA in the hopping regime regardless of the disorders in the dsDNA/dsRNA systems.

\subsection{Coherent Mechanism}

\begin{figure}[h!]
        \includegraphics[width=\columnwidth]{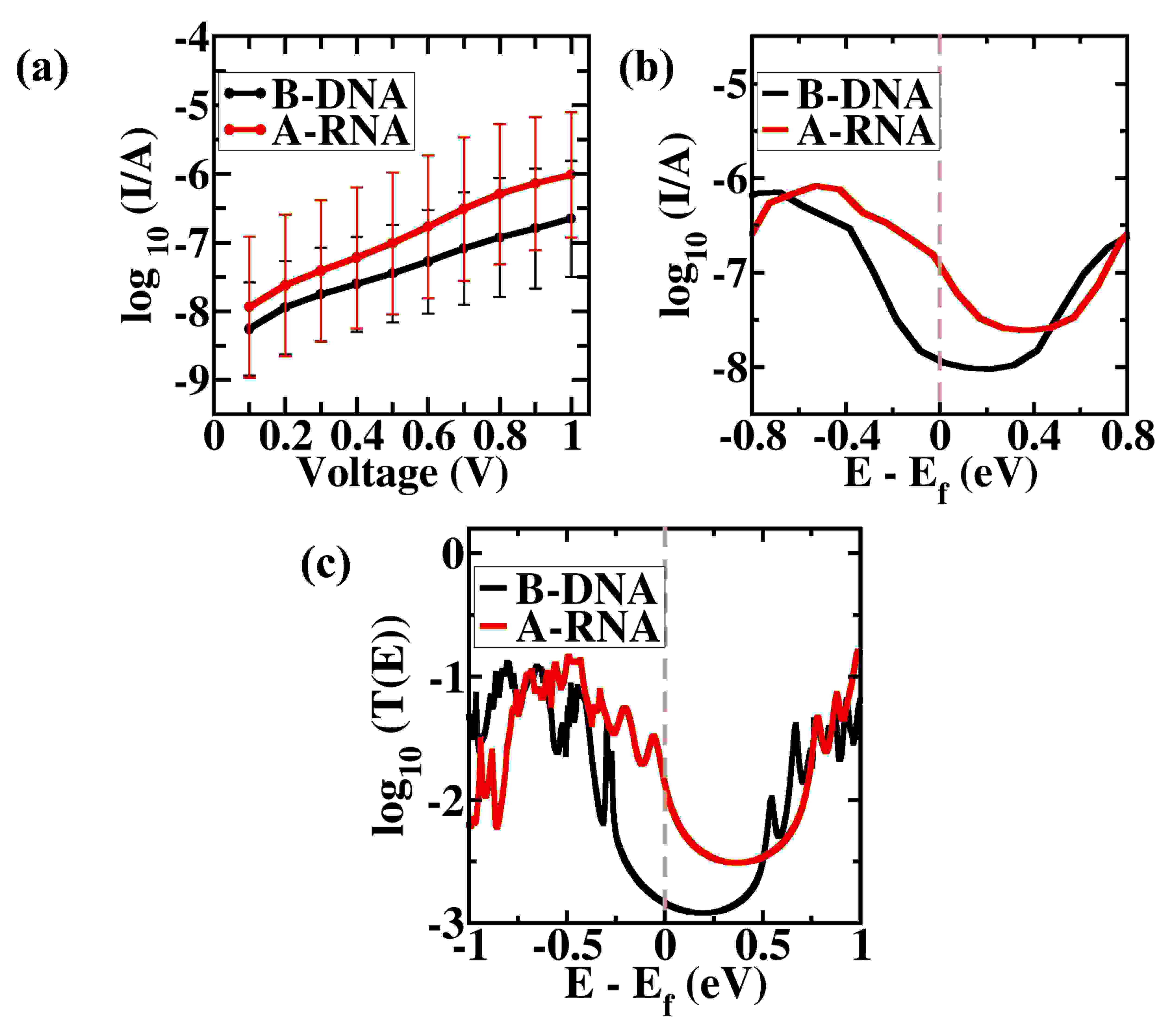}
        \caption{(a) V-I characteristics curve for 4 bp dsDNA vs dsRNA using coherent charge transport mechanism. Clearly, dsRNA conducts better than dsDNA. The error bars for the currents are derived assuming a log-normal distribution of the tunneling current. (b) Variation of current at 100 mV vs Fermi energy for 4 bp d-(CGCG) dsRNA and dsDNA. (c) Variation of transmission probability vs energy. The brown line represents the Fermi energy of the system. Both the curves show that at any energy, 4 bp dsRNA conducts better 4 bp dsDNA.} 
        \label{fig:fig_7}
\end{figure}

The V-I characteristics curves for 4 bp d-(CGCG) dsDNA and dsRNA sequences using coherent tunneling mechanism are shown in fig. \ref{fig:fig_7}(a).
Clearly, dsRNA has a higher conductance than dsDNA.
Notably, the current is of the order of few $\mu$A for both dsDNA and dsRNA.
This is in accordance with the results obtained using the hopping charge transport mechanism.

As the position of the Fermi level relative to the molecular orbital energies affects the magnitude of conductance in a drastic way, we sweep through multiple Fermi energies values between the molecular HOMO-LUMO gap and check the relative conductance of dsDNA and dsRNA.
Fig. \ref{fig:fig_7}(c) shows the logarithm of transmission probabilities vs energies for dsDNA and dsRNA showing that dsRNA has a higher transmission probability than dsDNA independent of the position of the Fermi energy.
This indicates that the current at 100 mV at any given Fermi energy will be higher for dsRNA.
This can be seen in fig. \ref{fig:fig_7}(b).
Hence, 4 bp dsRNA conducts better than 4 bp dsDNA.

\begin{figure}[h!]
        \includegraphics[width=\columnwidth]{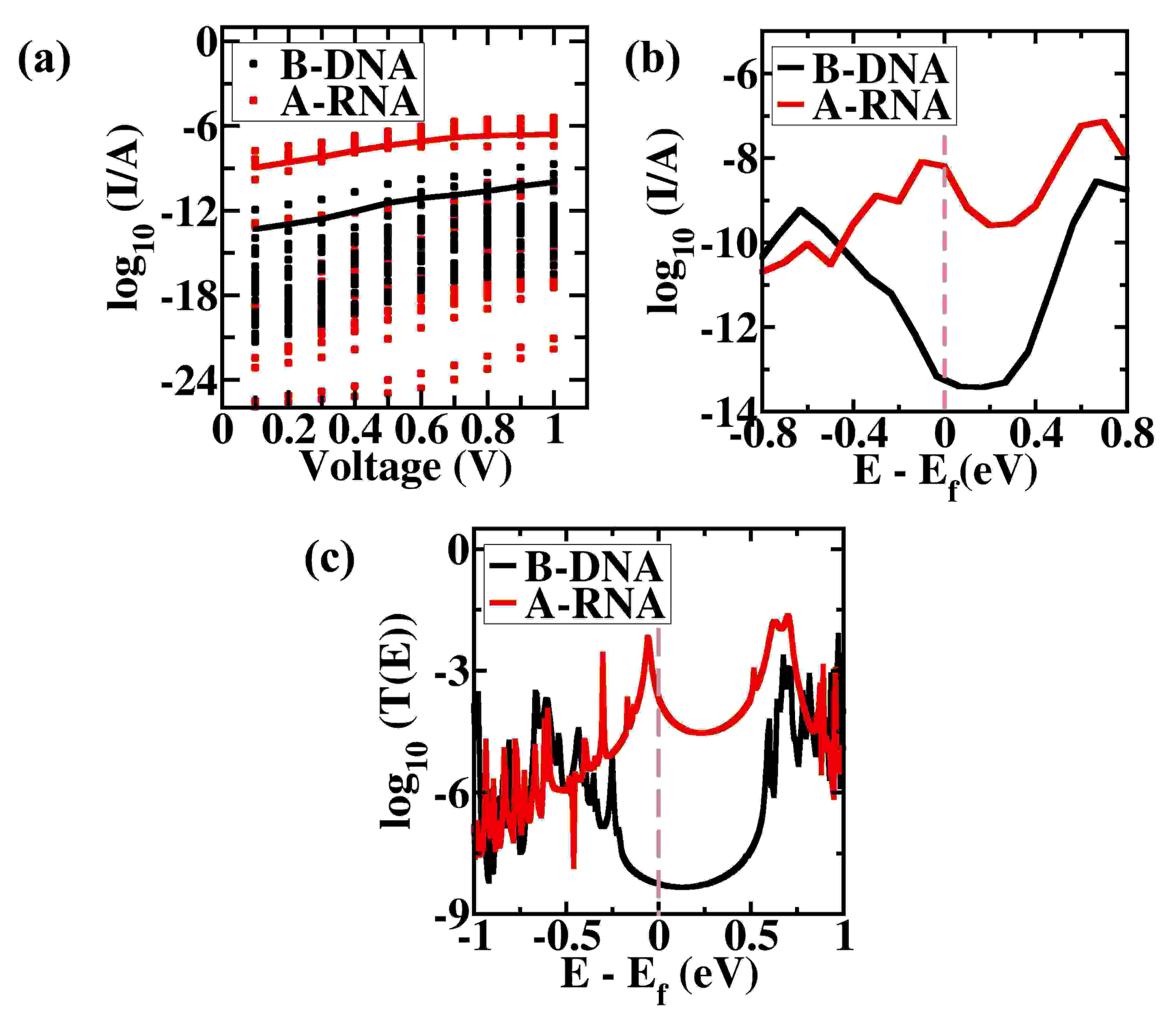}
        \caption{(a) V-I characteristics curve for 12 bp dsDNA d-(CGCGAATTCGCG) vs dsRNA d-(CGCGAAUUCGCG) using coherent charge transport mechanism. The solid lines represent the arithmetic means of the two nucleic acids. Considering arithmetic mean is acceptable here as the distributions are no longer expected to be log-normal. Clearly, dsRNA conducts better than dsDNA here. (b) Variation of current at 100 mV vs Fermi energy for 12 bp d-(CGCGAAUUCGCG) dsRNA and the corresponding dsDNA. (c) Variation of transmission probability vs energy. The brown line represents the Fermi energy of the system. Both the curves show that at any energy, 12 bp dsRNA conducts better 12 bp dsDNA.}
        \label{fig:fig_8}
\end{figure}

We also calculated transmission probabilities for 12 bp long dsDNA d-(CGCGAATTCGCG) and dsRNA d-(CGCGAAUUCGCG).
Fig. \ref{fig:fig_8}(a) shows the distribution of tunneling current of different snapshots of dsRNA and dsDNA at different potential biases.
The distribution of currents for dsDNA overlaps with a broad part of distribution of currents for dsRNA.
The solid lines represent the arithmetic means of the two nucleic acids.
Considering arithmetic mean is acceptable here as the distributions are no longer expected to be log-normal.
For longer dsDNA/dsRNA strands, a mixture of coherent mechanisms (tunneling as well as resonant transport) are operational leading to the observed distribution and averages\cite{venkatramani2014breaking}.
The current at 1 V is of the order of few nA in dsDNA which agrees with the experimental results\cite{artes2015conformational}.

Like 4 bp case, in 12 bp case also, at any given energy, the transmission probability of dsDNA is lower than that of dsRNA (fig. \ref{fig:fig_8}(c)).
Consequently, the current at 100 mV at any Fermi energy is also less for dsDNA than dsRNA (fig. \ref{fig:fig_8}(b)).

\begin{figure}[h!]
        \includegraphics[width=\columnwidth]{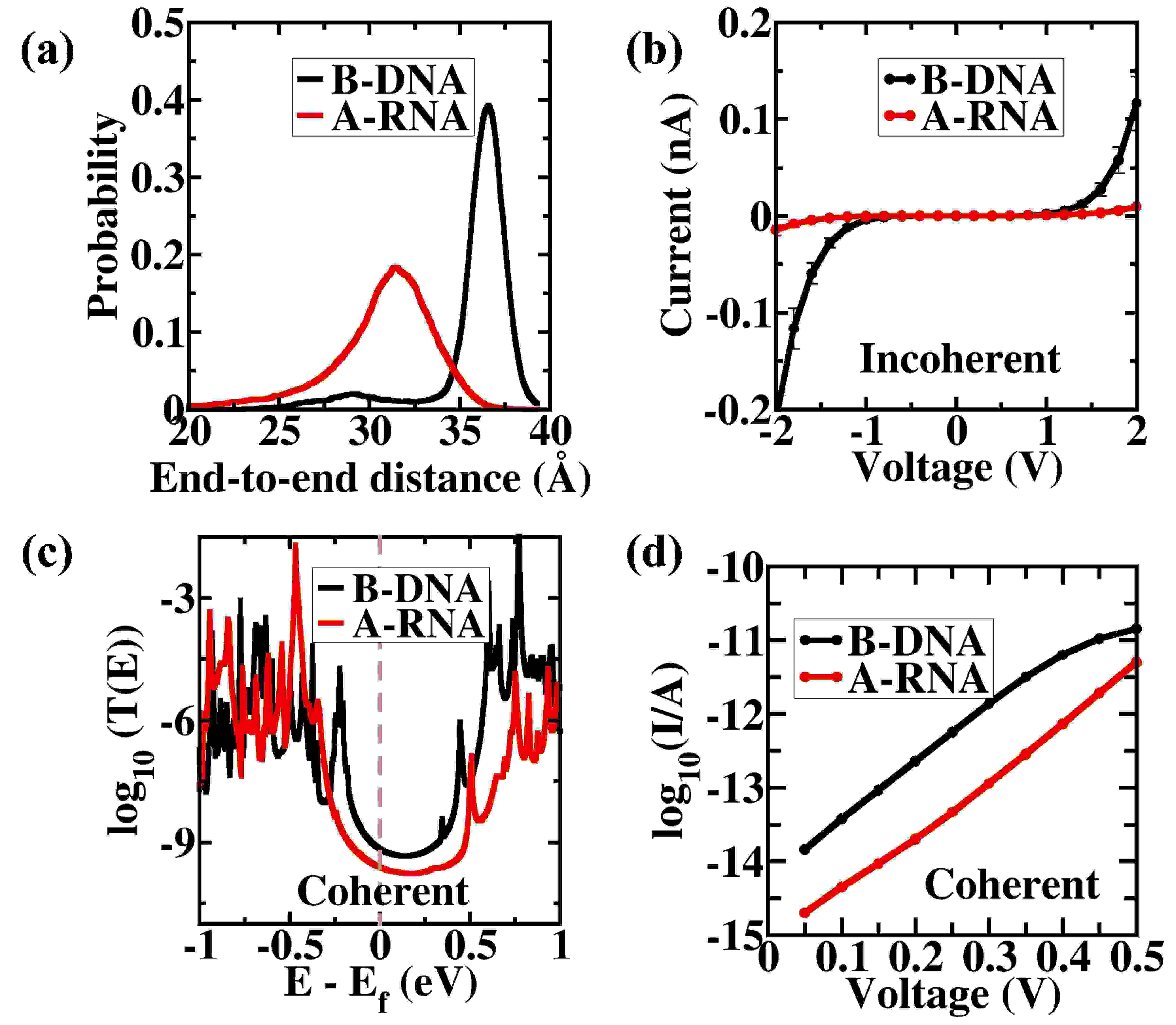}
	\caption{(a) Distribution of end-to-end distance of 12 bp A-RNA d-(CGCGAAUUCGCG) and B-DNA d-(CGCGAATTCGCG) for the 100 ns long simulation. B-DNA has a higher average end-to-end distance than A-RNA. (b) V-I characteristics for 20 structures of both 12 bp dsDNA and dsRNA having similar end-to-end distance using hopping charge transport mechanism. (c) Variation of transmission probability vs energy for the same 20 structures. (d) V-I characteristics for 20 structures of both 12 bp dsDNA and dsRNA having similar end-to-end distance using coherent tunneling charge transport mechanism. Clearly, in tunneling mechanism, dsDNA has higher transmission probabilities and coherent conductance relative to that of dsRNA when both have similar end-to-end lengths. In hopping case, B-DNA conducts better than A-RNA which shows that hopping phenomenon is more dependent on base-pairs rather than the end-to-end distance. }
    \label{fig:fig_9}
\end{figure}

The reason for this trend can be understood in terms of the high length dependence in tunneling mechanism\cite{hines2010transition}.
It is known that the conductance depends approximately exponentially on the length of the molecular device in tunneling mechanism\cite{hines2010transition,priyadarshy1998dna,magoga1998minimal}.
The end-to-end distance of A-form dsRNA is always less than B-form dsDNA for same base pair sequence.
For example, in the sequences used above, the distribution of end-to-end distance for 12 bp dsRNA shows that the average end-to-end distance for A-RNA is 30.97 \AA$ $ while for dsDNA, it is 38.72 \AA $ $  (fig. \ref{fig:fig_9}(a)).
Due to this, dsRNA conducts better than dsDNA for a given Fermi energy level.
For 4 bp sequence, the difference in the end-to-end distance is lower which leads to less difference in the order of magnitude of conductance.

	Since the tunneling mechanism is highly length dependent phenomenon\cite{hines2010transition}, comparing the B-DNA and A-RNA sequences with similar end-to-end distances should provide a better insight to their physical properties.
For this purpose, we choose 20 snapshots both of 12 bp A-RNA as well as of 12 bp B-DNA which have similar end-to-end distance (between 34.6 \AA$ $ and 34.7 \AA$ $ (fig. \ref{fig:fig_9}(a))) and calculate both hopping and tunneling charge transport.
We find that, dsDNA shows better conductance than dsRNA for both incoherent hopping transport (fig. \ref{fig:fig_9}(b)) as well as coherent tunneling transport(fig. \ref{fig:fig_9}(c) and \ref{fig:fig_9}(d)).
However, in a given ensemble of B-DNA and A-RNA strutures of the same base pair sequences, A-RNA should be more compact, on average, and therefore show higher coherent conductance relative to B-DNA (fig. \ref{fig:fig_8}(a)).
This is consistent with our conclusion above that the higher conductance of dsRNA in the tunneling regime is due to its shorter end-to-end distance relative to dsDNA since the structures having similar length show similar relative conductance as that obtained in the hopping regime.
The V-I characteristics of the other 12 bp dsDNA and dsRNA sequences calculated using tunneling mechanism also show that dsRNA conducts better than dsDNA because of the smaller end-to-end length of dsRNA relative to dsDNA (fig. S3 of ESI).

 The hopping mechanism is a weakly length-dependent process\cite{hines2010transition} and the electronic coupling between two charge hopping sites plays a vital role in determining the charge transport properties.
 Hence, factors such as the local disorders and high flexibility of nucleobase stacking reduce the conductance considerably in the hopping mechanism.
For hopping transport, the local structural parameters like rise, slide and twist affect the conductance more than the end-to-end distance.
Hence, in the hopping regime, B-DNA conducts better than A-RNA due to the above reason.
 In tunneling mechanism, at long distances, the charge transfer drops approximately exponentially with distance\cite{priyadarshy1998dna} and hence the end-to-end length of the nucleic acid affects the conductance most.
The decay constant values for dsDNA and dsRNA are 0.52 \AA$^{-1}$ and 0.74 \AA$^{-1}$ respectively which represents strong length dependence in coherent tunneling regime (fig. S7b in ESI).
Also, dsRNA has higher decay constant relative to dsDNA suggesting stronger coherent mechanisms in dsRNA.
This is in accordance with previous experimental results \cite{li2016comparing,artes2015conformational}.

\begin{figure}[h!]
        \includegraphics[width=\columnwidth]{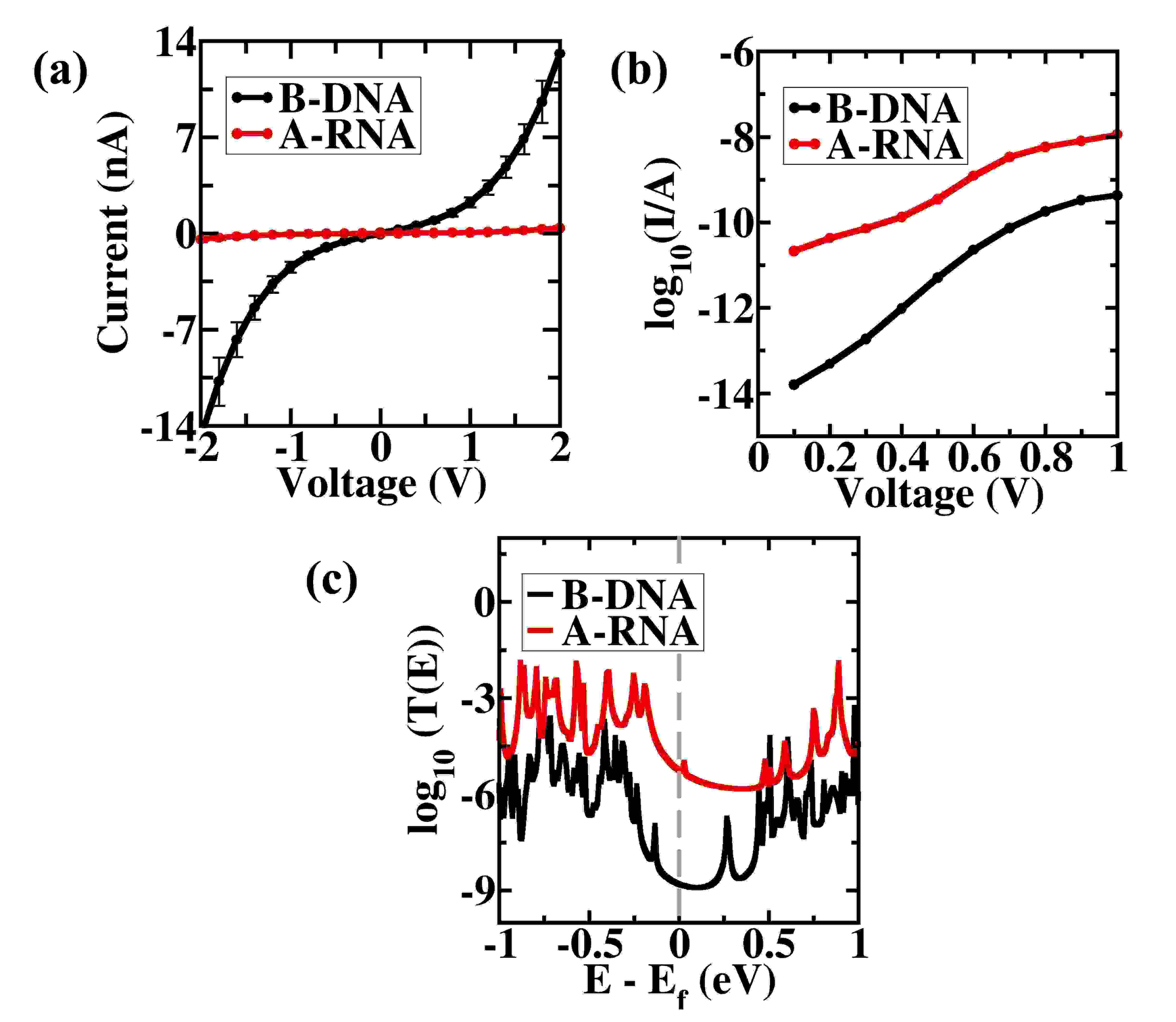}
        \caption{(a) V-I characteristics curve for 9 bp (d-(CCCGCGCCC)) dsDNA vs dsRNA using hopping charge transport mechanism. The charge transport properties of this sequence have been studied experimentally in Ref. \citenum{artes2015conformational}. dsRNA has almost two orders of magnitude less conductance than dsDNA. (b) V-I characteristics of both 9 bp (d-(CCCGCGCCC)) dsDNA and dsRNA using coherent tunneling charge transport mechanism. (c) Variation of transmission probability vs energy for 9 bp dsDNA and dsRNA. Clearly, in the tunneling mechanism, dsRNA conducts better than dsDNA by an order of magnitude which is seen experimentally as well.}
	\label{fig:fig_10}
\end{figure}

In order to compare our results with available experiments, we also calculated the charge transport properties of 9 bp B-form dsDNA and A-form dsRNA with sequence d-(CCCGCGCCC).
The charge transport properties of this sequence have been studied experimentally in Ref. \citenum{artes2015conformational}.
The A-form 9-mer DNA:RNA hybrid used in Ref. \citenum{artes2015conformational} has almost one order of magnitude higher conductance than B-form 9-mer  dsDNA.
In our study, we find a similar result for tunneling transport (fig. \ref{fig:fig_10}(b) and fig. \ref{fig:fig_10}(c)).
On the other hand, for hopping transport (fig. \ref{fig:fig_10}(a)), although the current is in nA range which is consistent with the experimentally observed current range\cite{artes2015conformational}, B-form dsDNA has almost 2 orders of magnitude higher conductance than A-form dsRNA.
We also calculated the charge transport properties of 13 bp B-form dsDNA and A-form dsRNA with sequence d-(CCCGCGCGCGCCC) which also shows similar trends and the results are shown in fig. S6.
 Hihath and co-workers\cite{artes2015conformational,li2016comparing, li2016long} find A-form DNA:RNA hybrids to be around 10 times more conductive than B-form dsDNA showing the conformational gating of dsDNA conductance in their experimental works. 
 Moreover, they obtain a higher decay constant for DNA:RNA hybrids relative to dsDNA. 
 Our simulations also show  higher electronic conduction and decay constant for dsRNA relative to dsDNA in the tunneling regime, generalizing the conformational gating behavior of nucleic acid conductance.
 In another series of experimental works, Tao and co-workers\cite{xiang2015intermediate, li2016thermoelectric, xu2004direct, xiang2017gate} study the length and sequence dependence of dsDNA conductance and find dsDNA resistance to be of the order of M$\Omega$ to G$\Omega$ which is in close agreement to the order of magnitude of resistance observed from our calculations.

\section{Conclusions}

We have calculated and compared the charge transport properties of dsDNA and dsRNA in two different charge transport regimes.
In the diffusive limit, i.e. the hopping mechanism, dsDNA conducts better than dsRNA as the higher values of slide and inclination angle in dsRNA lead to lower currents in dsRNA despite its lower helical rise.
In the coherent limit as well, i.e. wherein a tunneling mechanism is dominant, dsDNA conducts better than dsRNA sequences of similar lengths but because of the smaller average end-to-end distance of dsRNA, it will show better conductance than dsDNA for a general ensemble of structures.
Thus, the answer to the question 'DNA or RNA- which one conducts better?' depends on the regime under which the molecular charge transport is measured.
In the incoherent hopping regime, dsDNA has higher conductance relative to dsRNA as it is more ordered than dsRNA whereas in the coherent regime, the observed conductance trend is reversed as dsRNA is shorter than dsDNA.
By knowing the regime in which the molecule is conducting, one can easily get a fair estimate of its electrical properties.
Nonetheless, higher flexibility and comparable conductance efficiencies of dsRNA relative to dsDNA make dsRNA also suitable for all the organic electronics applications.
This work provides an understanding of the charge transport phenomenon involved in the dsDNA and dsRNA which can contribute significantly towards the field of molecular electronics and RNA nanotechnology.

\section*{Conflicts of interest}
There are no conflicts to declare.

\section*{Acknowledgements}
We thank Prof. Thomas E. Cheatham, III for his insightful remarks. 
We thank Mr. Ravinder Kumar for helpful discussions.
We also thank Dr. Veerabhadrarao Kaliginedi for his useful comments.
We acknowledge financial support from DST, India.  
We thank TUE-CMS, IISc for the computation time. 
A.A. thanks MHRD, India for the generous fellowship.

\bibliography{references} 
\bibliographystyle{rsc} 

\end{document}


\title{DNA versus RNA - which shows higher electronic conduction?: Supporting Information}
\author{Abhishek Aggarwal}
\affiliation{Center for Condensed Matter Theory, Department of Physics, Indian Institute of Science, Bangalore-560012, India}
\author{Saientan Bag}
\affiliation{Center for Condensed Matter Theory, Department of Physics, Indian Institute of Science, Bangalore-560012, India}
\author{Ravindra Venkatramani}
\affiliation{Department of Chemical Sciences, Tata Institute of Fundamental Research, Colaba, Mumbai-400005, India}
\author{Manish Jain}
\affiliation{Center for Condensed Matter Theory, Department of Physics, Indian Institute of Science, Bangalore-560012, India}
\author{Prabal K. Maiti}
\email{maiti@iisc.ac.in}
\affiliation{Center for Condensed Matter Theory, Department of Physics, Indian Institute of Science, Bangalore-560012, India}

\pacs{}

\maketitle

\section*{ Hole Current for different 12 bp seqeunces of dsDNA and dsRNA}

 \begin{figure}[!htb]
    \centering
    \begin{subfigure}[t]{0.4\columnwidth}
        \centering
        \includegraphics[width=\columnwidth]{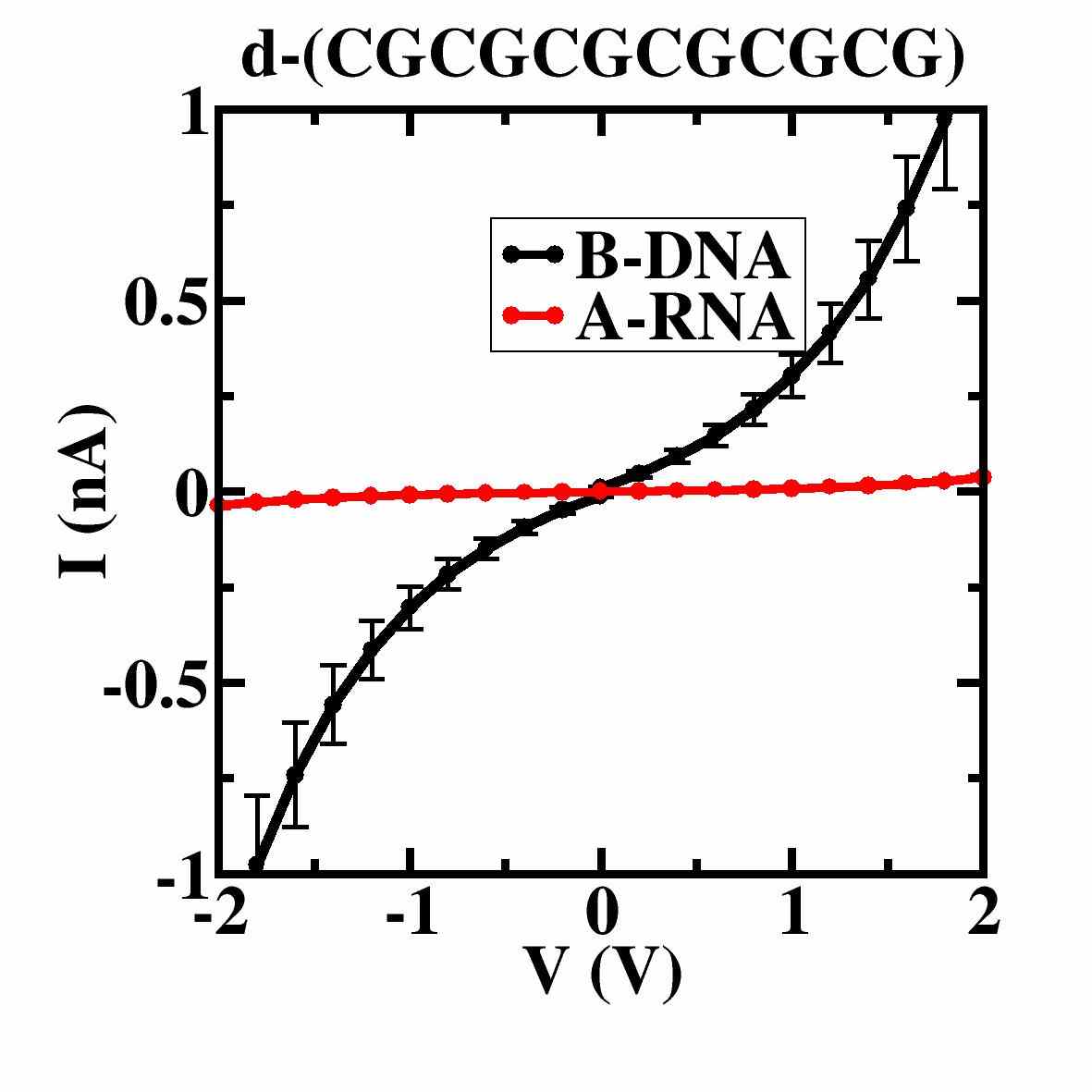}
\caption{}
    \end{subfigure}%
   \hfill
    \begin{subfigure}[t]{0.4\columnwidth}
        \centering
        \includegraphics[width=\columnwidth]{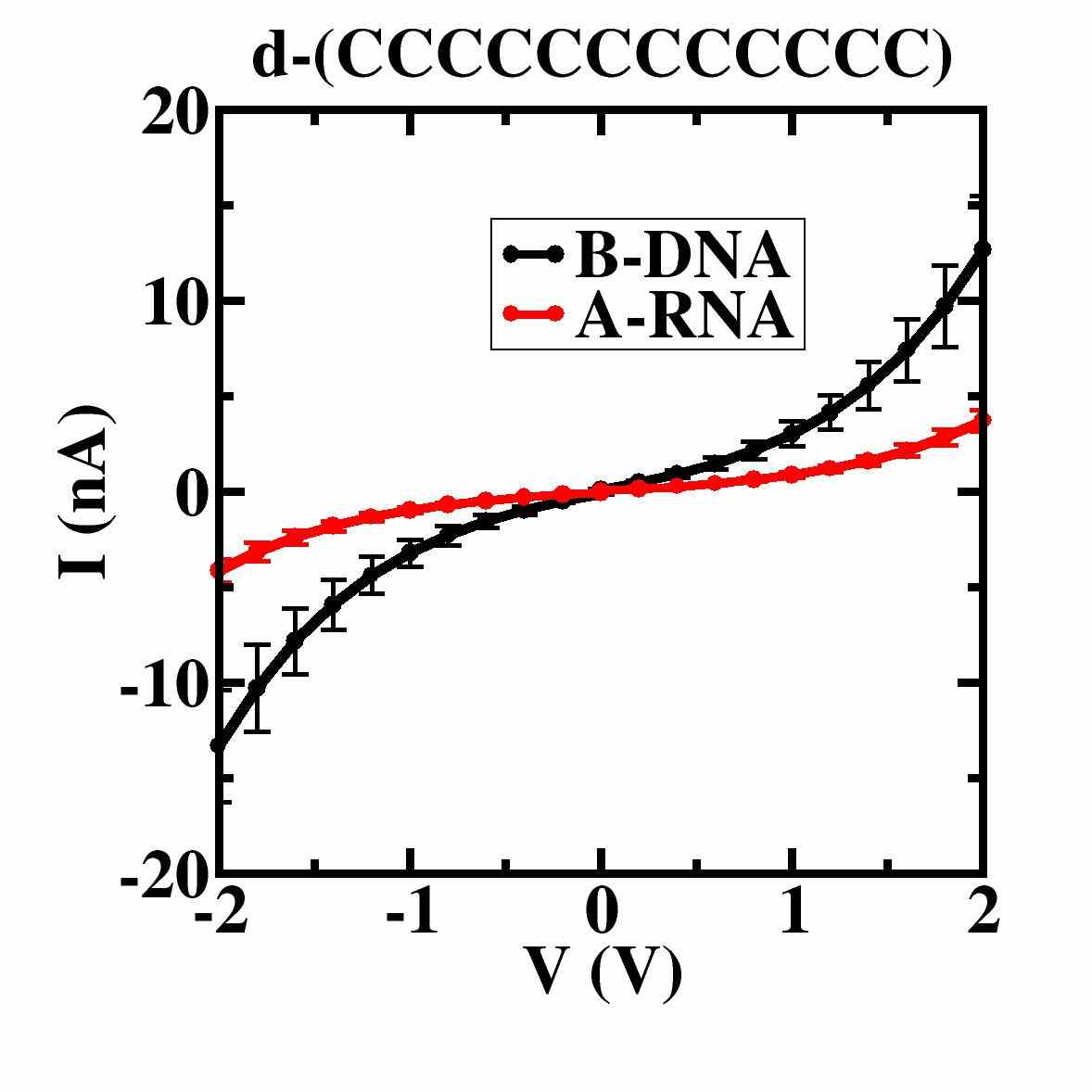}
\caption{}
\end{subfigure}

    \centering
    \begin{subfigure}[t]{0.4\columnwidth}
        \centering
        \includegraphics[width=\columnwidth]{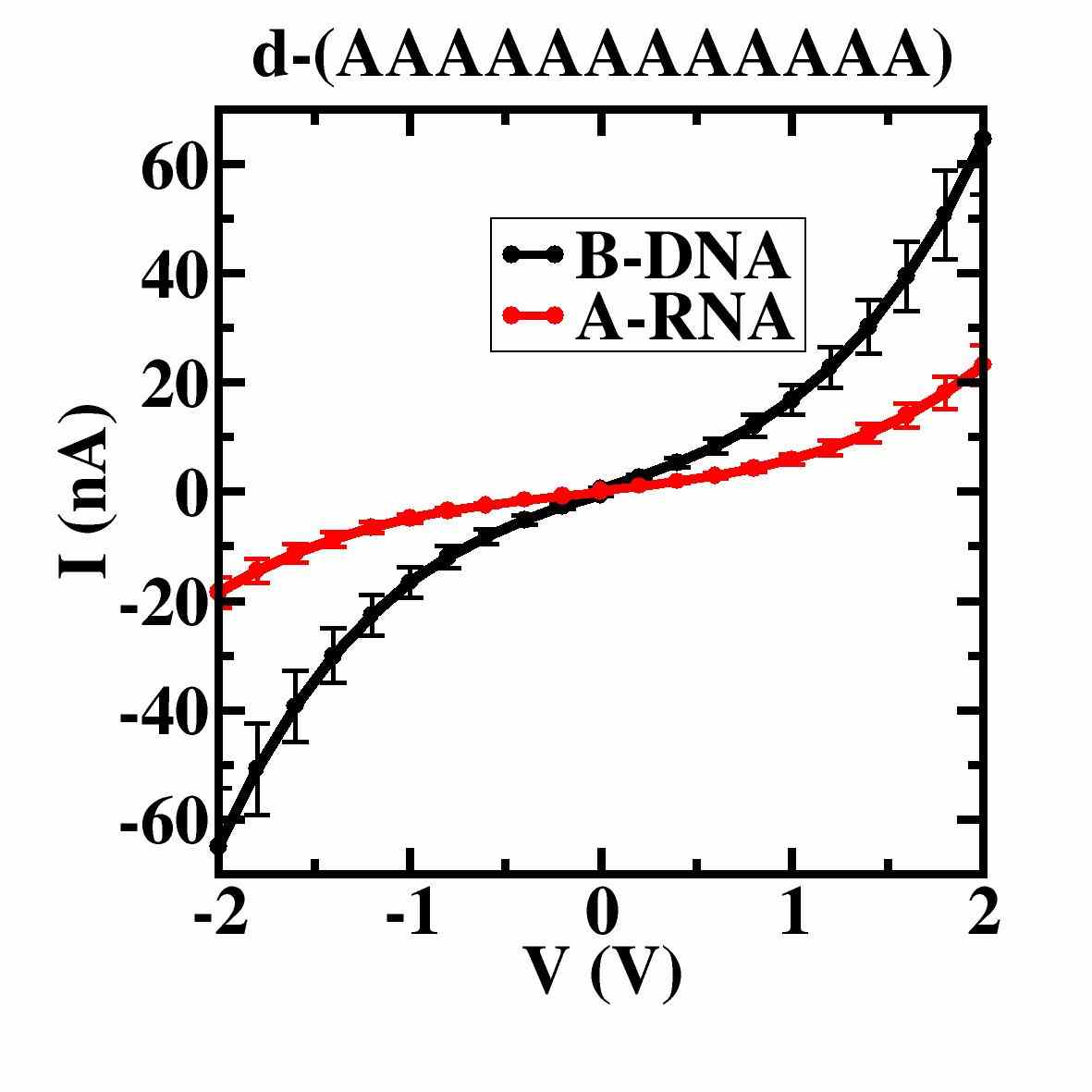}
\caption{}
    \end{subfigure}%
   \hfill
    \begin{subfigure}[t]{0.4\columnwidth}
        \centering
        \includegraphics[width=\columnwidth]{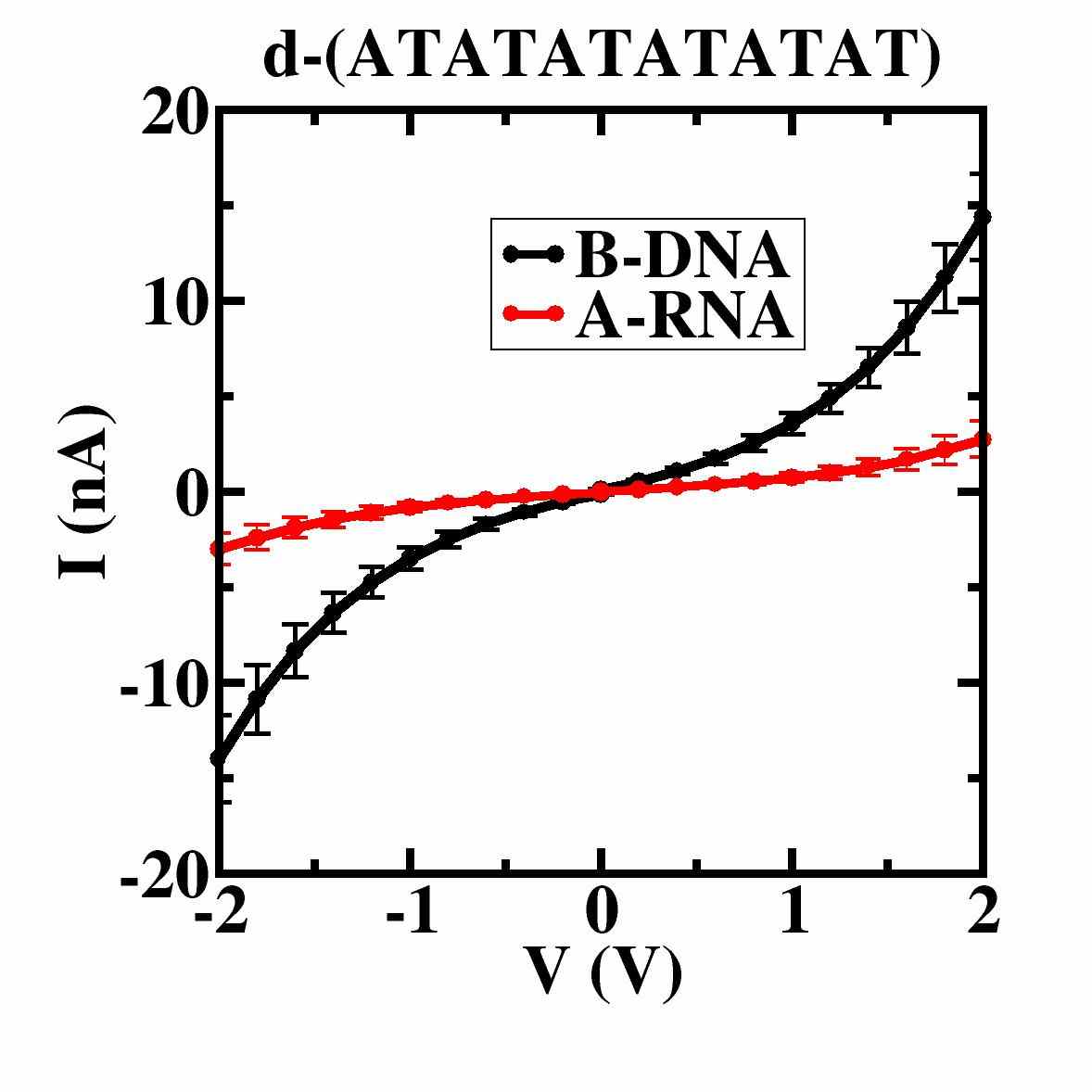}
\caption{}
\label{fig:AT_hopping}
\end{subfigure}

\captionsetup{justification=raggedright, singlelinecheck=false,font=footnotesize}
	\caption{Hole current for (a) $d-(CG)_{6}$ (b) $d-(CC)_{6}$ (c) $d-(AA)_{6}$ (d) $d-(AT)_{6}$ sequences of dsDNA and dsRNA using hopping mechanism. For each sequence dsDNA conducts better than dsRNA in incoherent regime. }
    \label{fig:same_sequence_hole}
\end{figure}

Hole current for different sequences of dsDNA and dsRNA using hopping mechanism are shown in fig. \ref{fig:same_sequence_hole}. For each sequence, dsDNA conducts better than dsRNA in incoherent regime.

\clearpage

\section{ Electronic Current for different 12 bp seqeunces of dsDNA and dsRNA}

 \begin{figure}[!htb]
    \centering
    \begin{subfigure}[t]{0.4\columnwidth}
        \centering
        \includegraphics[width=\columnwidth]{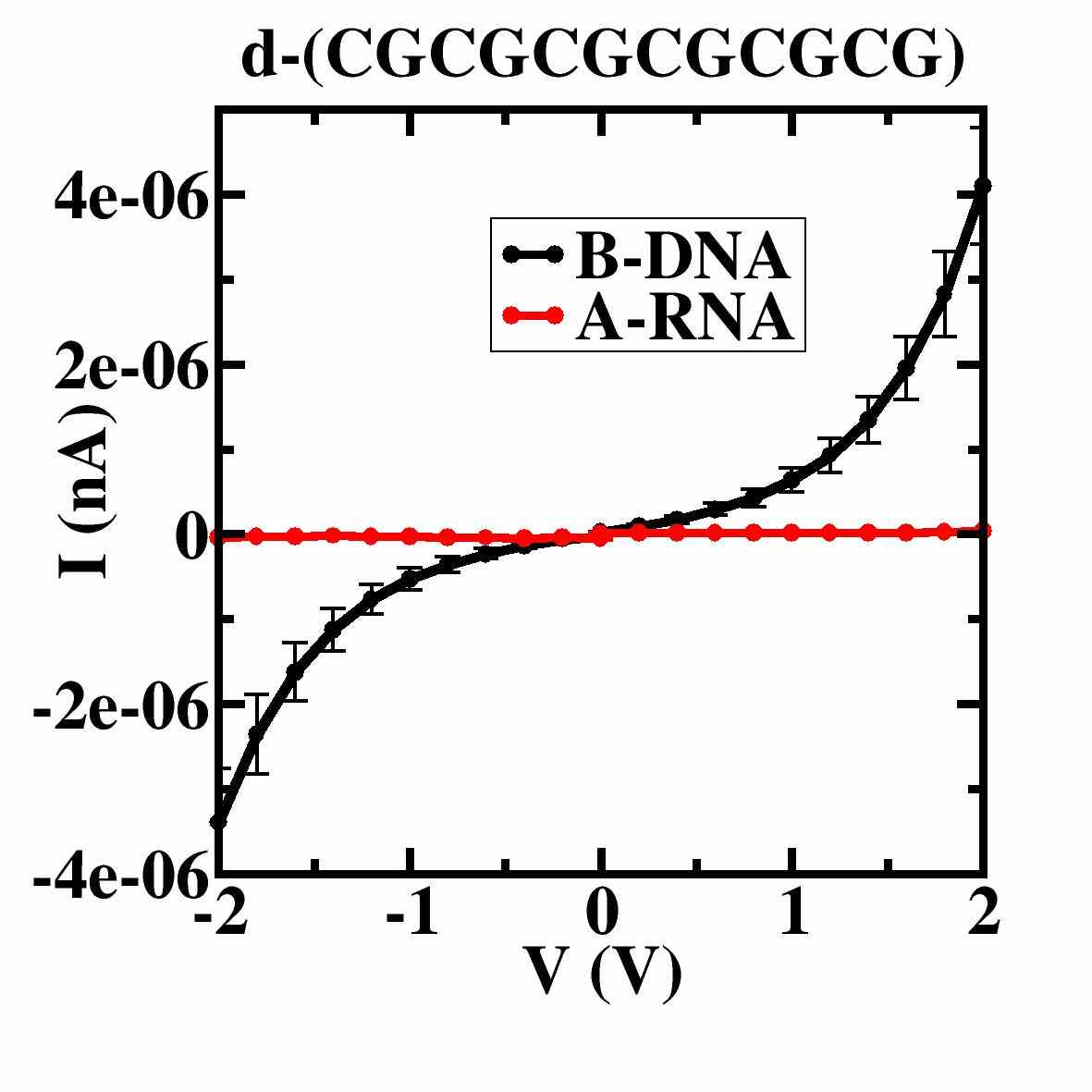}
\caption{}
    \end{subfigure}%
   \hfill
    \begin{subfigure}[t]{0.4\columnwidth}
        \centering
        \includegraphics[width=\columnwidth]{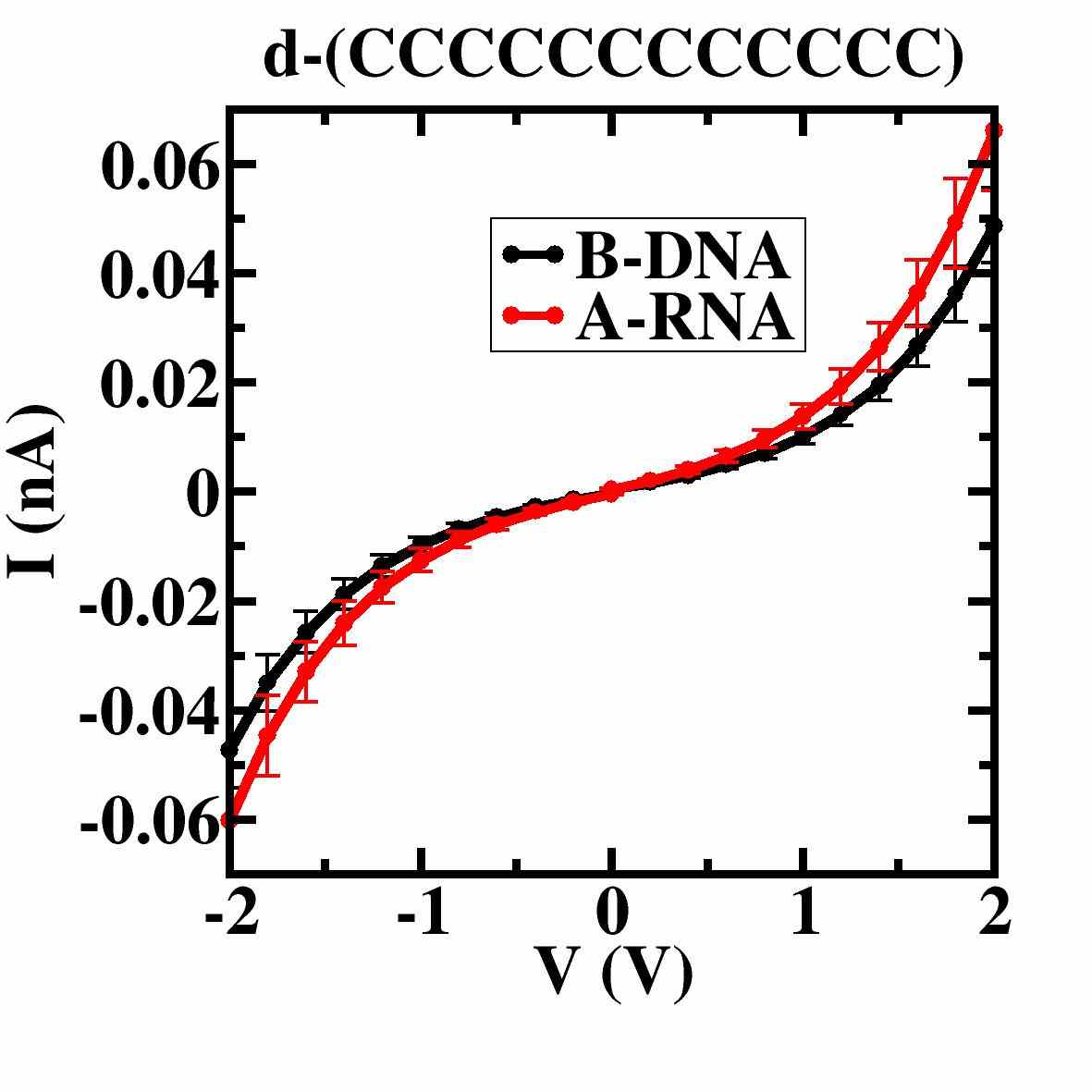}
\caption{}
\end{subfigure}
    \centering
    \begin{subfigure}[t]{0.4\columnwidth}
        \centering
        \includegraphics[width=\columnwidth]{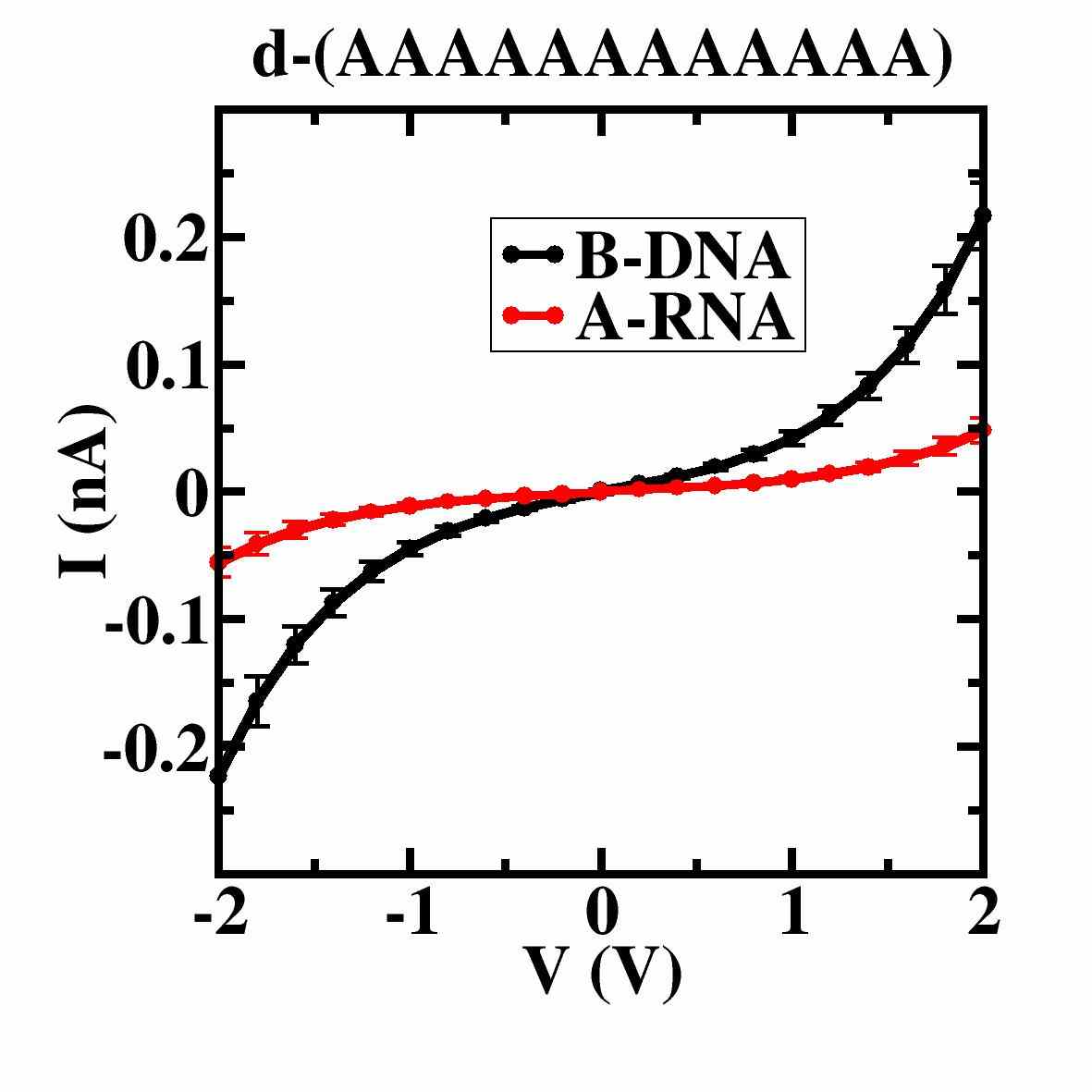}
\caption{}
    \end{subfigure}%
   \hfill
    \begin{subfigure}[t]{0.4\columnwidth}
        \centering
        \includegraphics[width=\columnwidth]{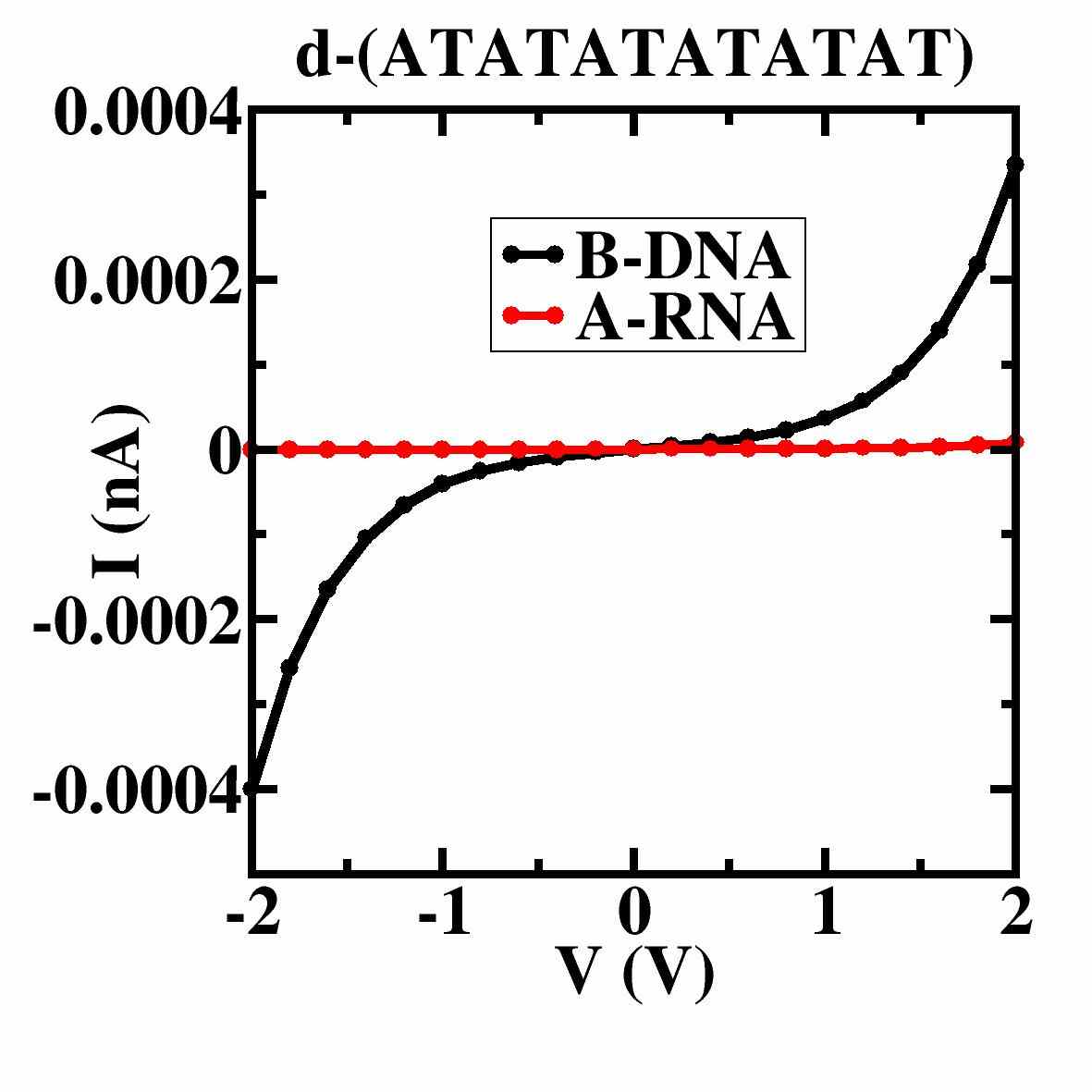}
\caption{}
\label{fig:AT_hopping}
\end{subfigure}
\captionsetup{justification=raggedright, singlelinecheck=false,font=footnotesize}
        \caption{Electronic current for (a) $d-(CG)_{6}$ (b) $d-(CC)_{6}$ (c) $d-(AA)_{6}$ (d) $d-(AT)_{6}$ sequences of dsDNA and dsRNA using hopping mechanism. For each sequence dsDNA conducts better than dsRNA in incoherent regime. }
    \label{fig:same_sequence_elec}
\end{figure}

Electronic current for different sequences of dsDNA and dsRNA using hopping mechanism are shown in fig. \ref{fig:same_sequence_elec}.
Although, the electronic current is lower in magnitude than hole current, dsDNA conducts better than dsRNA for each sequence in incoherent regime.

\clearpage

\section{Current for different 12 bp seqeunces of dsDNA and dsRNA using coherent tunneling mechanism}

 \begin{figure}[!htb]
    \centering
    \begin{subfigure}[t]{0.23\columnwidth}
        \centering
        \includegraphics[width=\columnwidth]{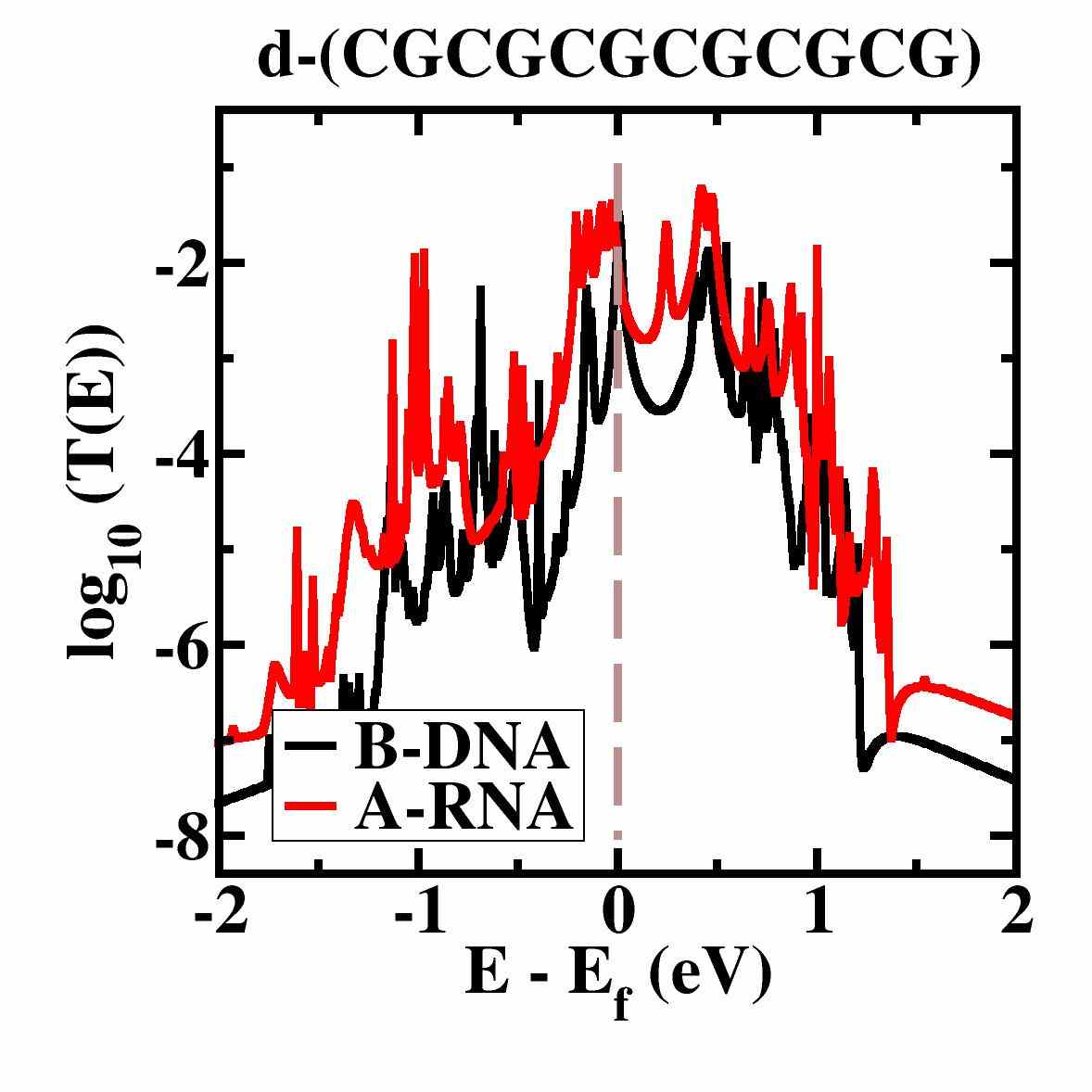}
\caption{}
    \end{subfigure}%
   \hfill
    \begin{subfigure}[t]{0.23\columnwidth}
        \centering
        \includegraphics[width=\columnwidth]{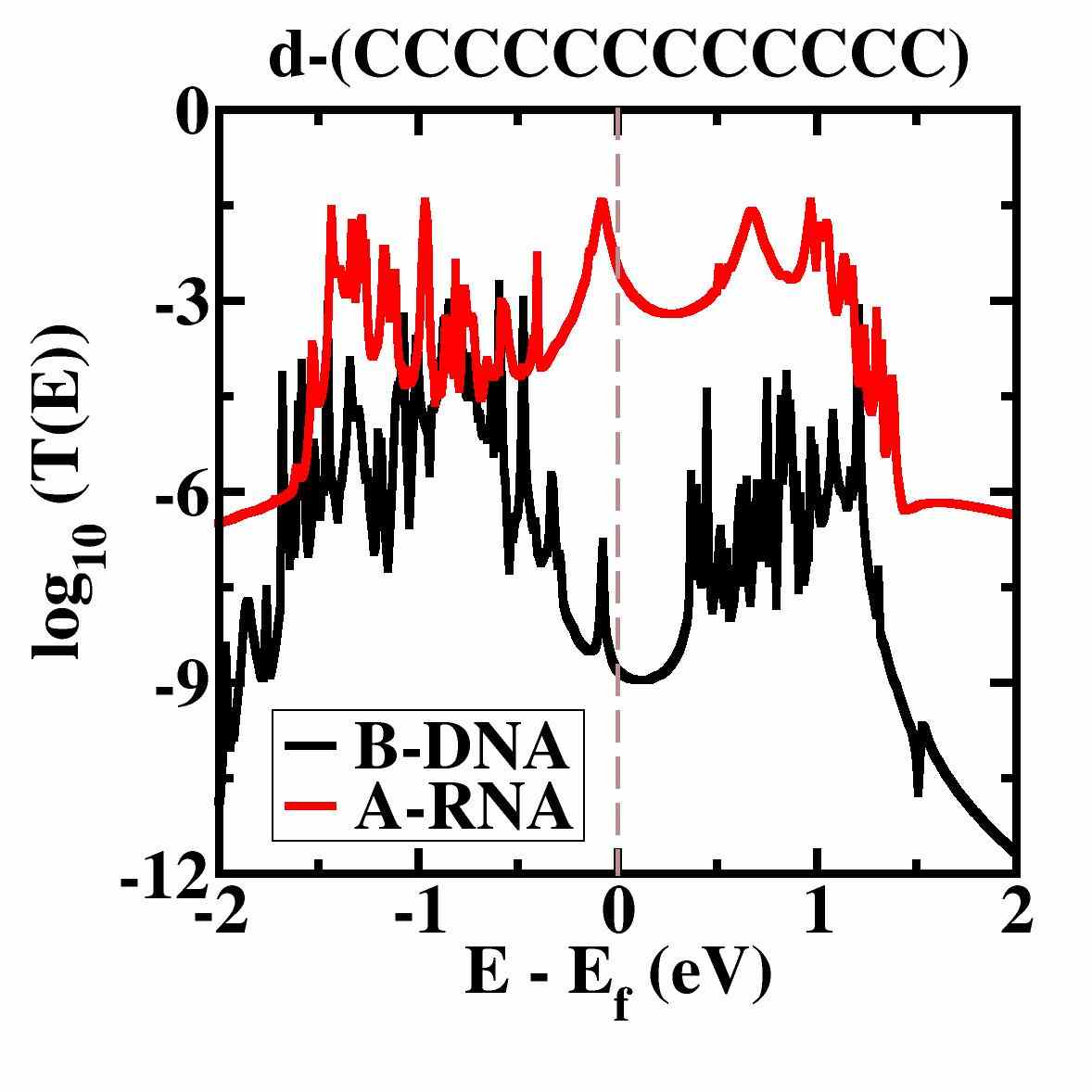}
\caption{}
\end{subfigure}
    \centering
    \begin{subfigure}[t]{0.23\columnwidth}
        \centering
        \includegraphics[width=\columnwidth]{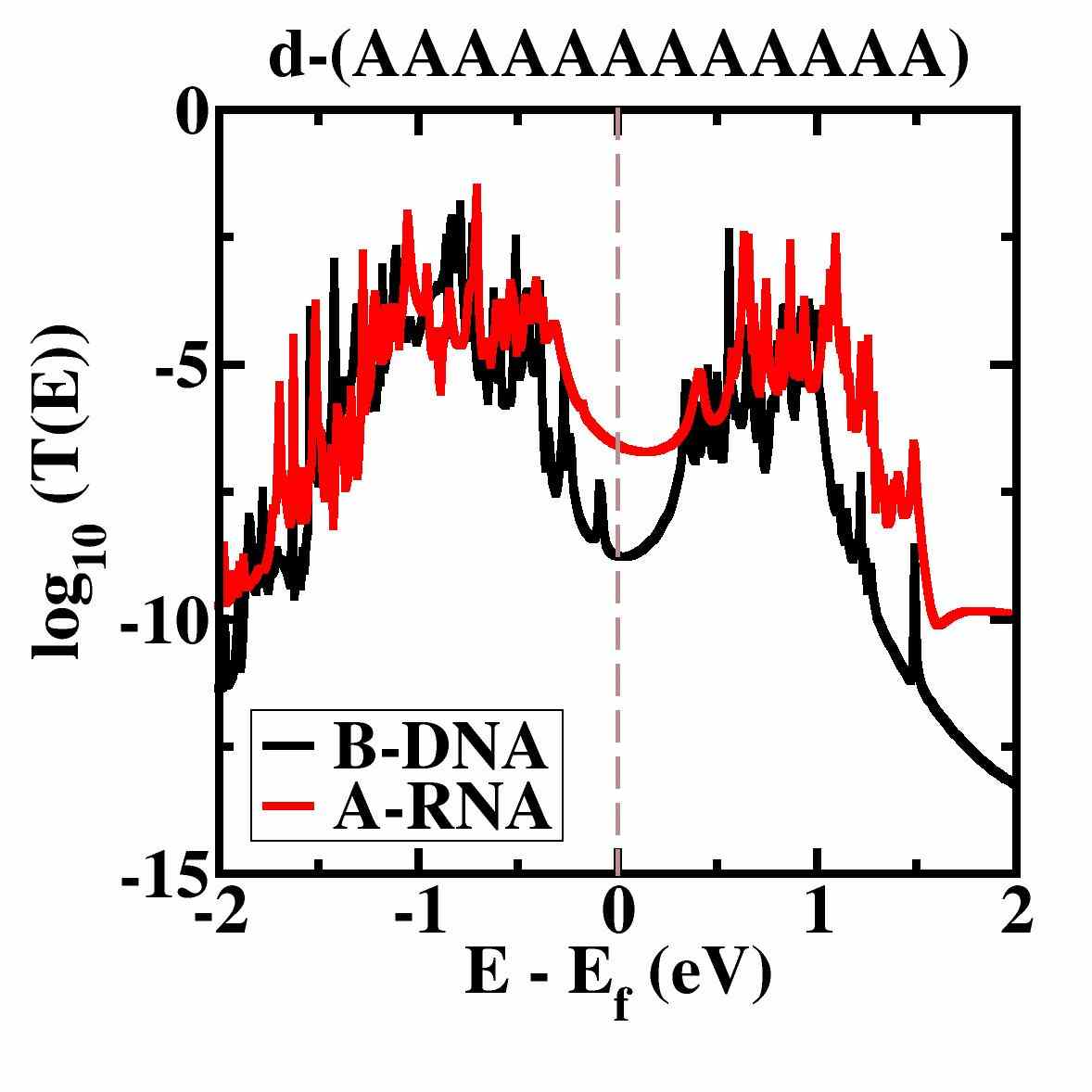}
\caption{}
    \end{subfigure}%
   \hfill
    \begin{subfigure}[t]{0.23\columnwidth}
        \centering
        \includegraphics[width=\columnwidth]{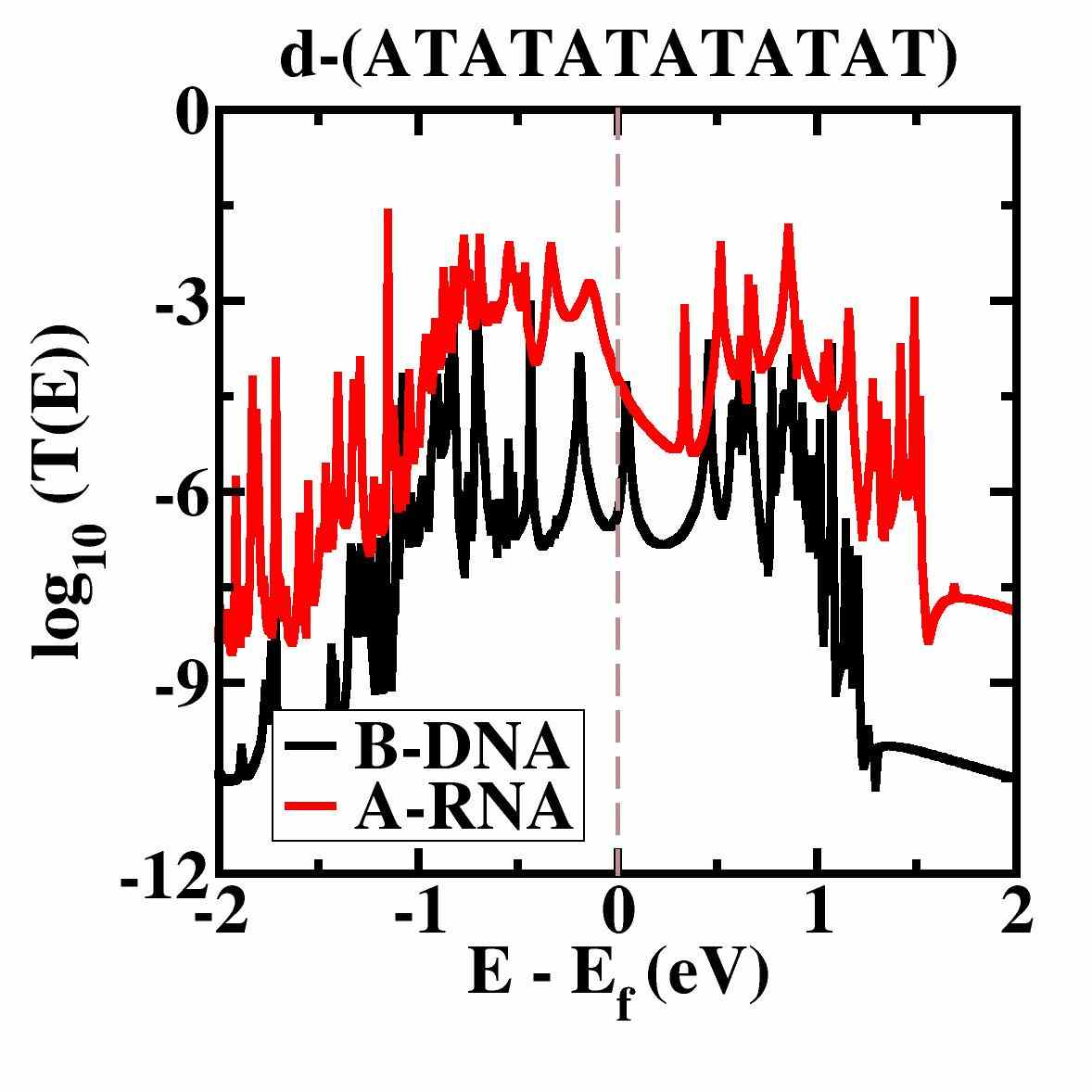}
\caption{}
\label{fig:AT_trans}
\end{subfigure}
\captionsetup{justification=raggedright, singlelinecheck=false,font=footnotesize}
	 \caption{Transmission coefficient near Fermi energy region for (a) d-(CGCGCGCGCGCG) (b) d-(CCCCCCCCCCCC) (c) d-(AAAAAAAAAAAA) (d) d-(ATATATATATAT) sequences of dsDNA and dsRNA using coherent tunneling mechanism. For each sequence dsRNA has higher transmission probability than dsDNA in coherent regime. }
    \label{fig:same_sequence_tunnel}
\end{figure}

\begin{figure}[!htb]
    \centering
    \begin{subfigure}[t]{0.23\columnwidth}
        \centering
        \includegraphics[width=\columnwidth]{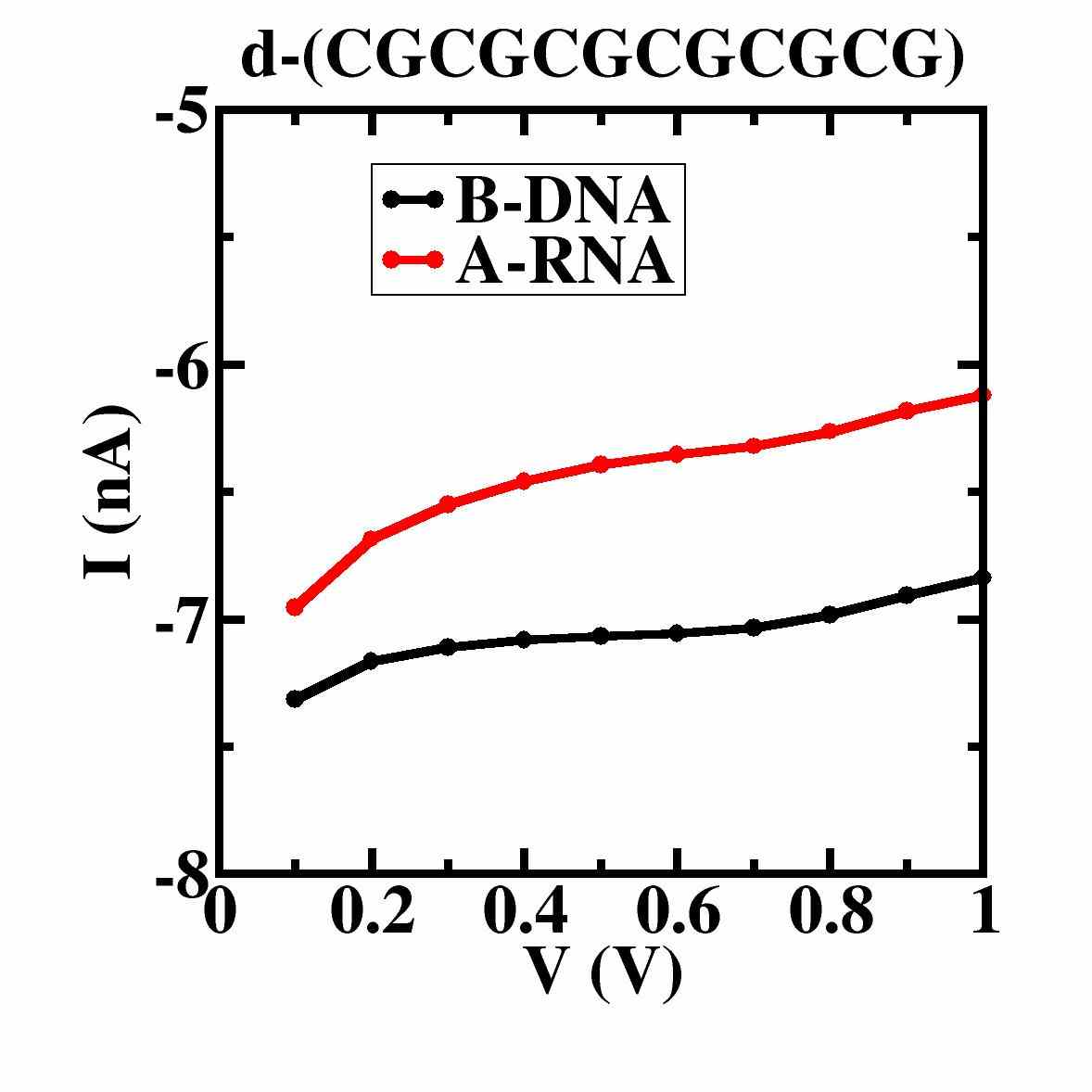}
\caption{}
    \end{subfigure}%
   \hfill
    \begin{subfigure}[t]{0.23\columnwidth}
        \centering
        \includegraphics[width=\columnwidth]{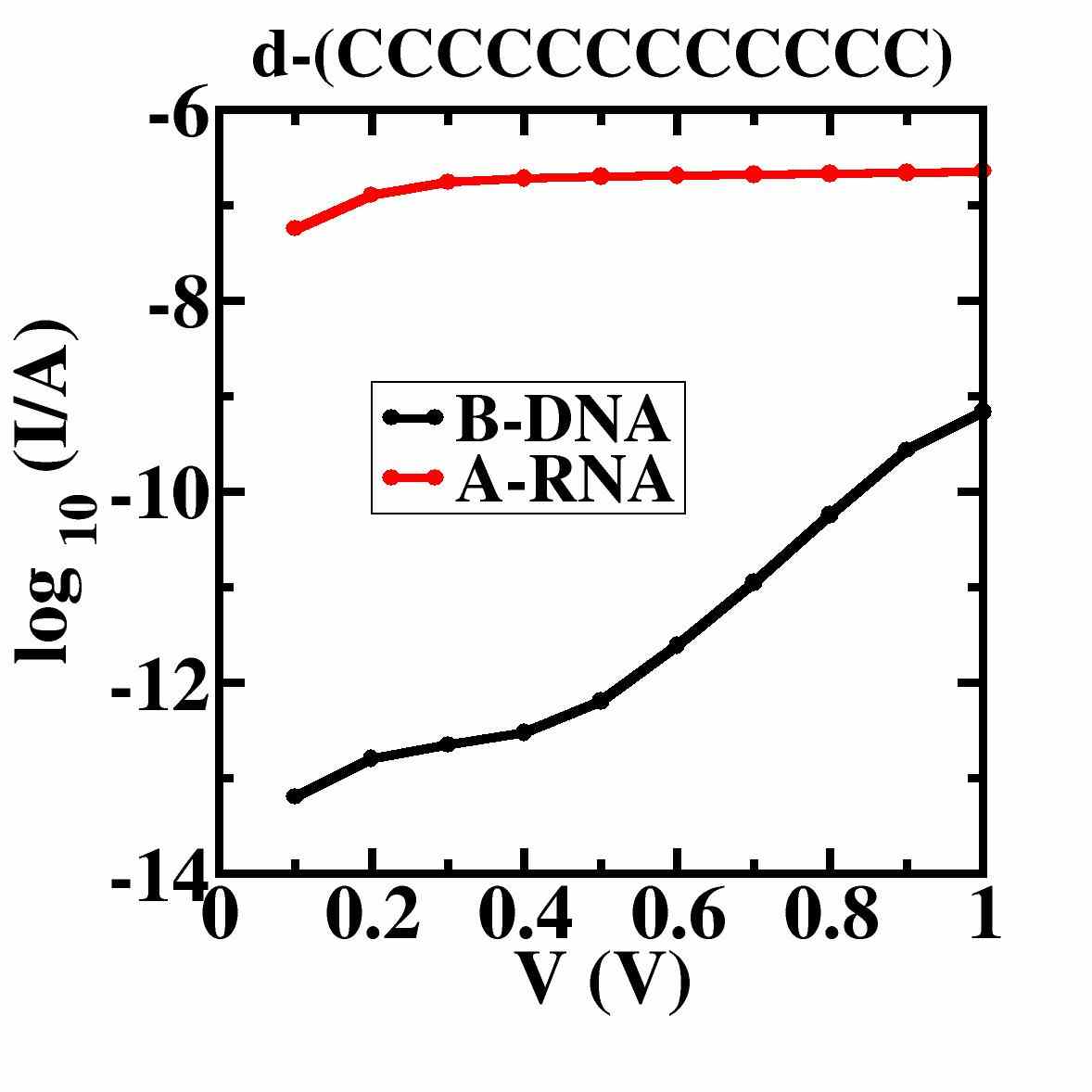}
\caption{}
\end{subfigure}
    \centering
    \begin{subfigure}[t]{0.23\columnwidth}
        \centering
        \includegraphics[width=\columnwidth]{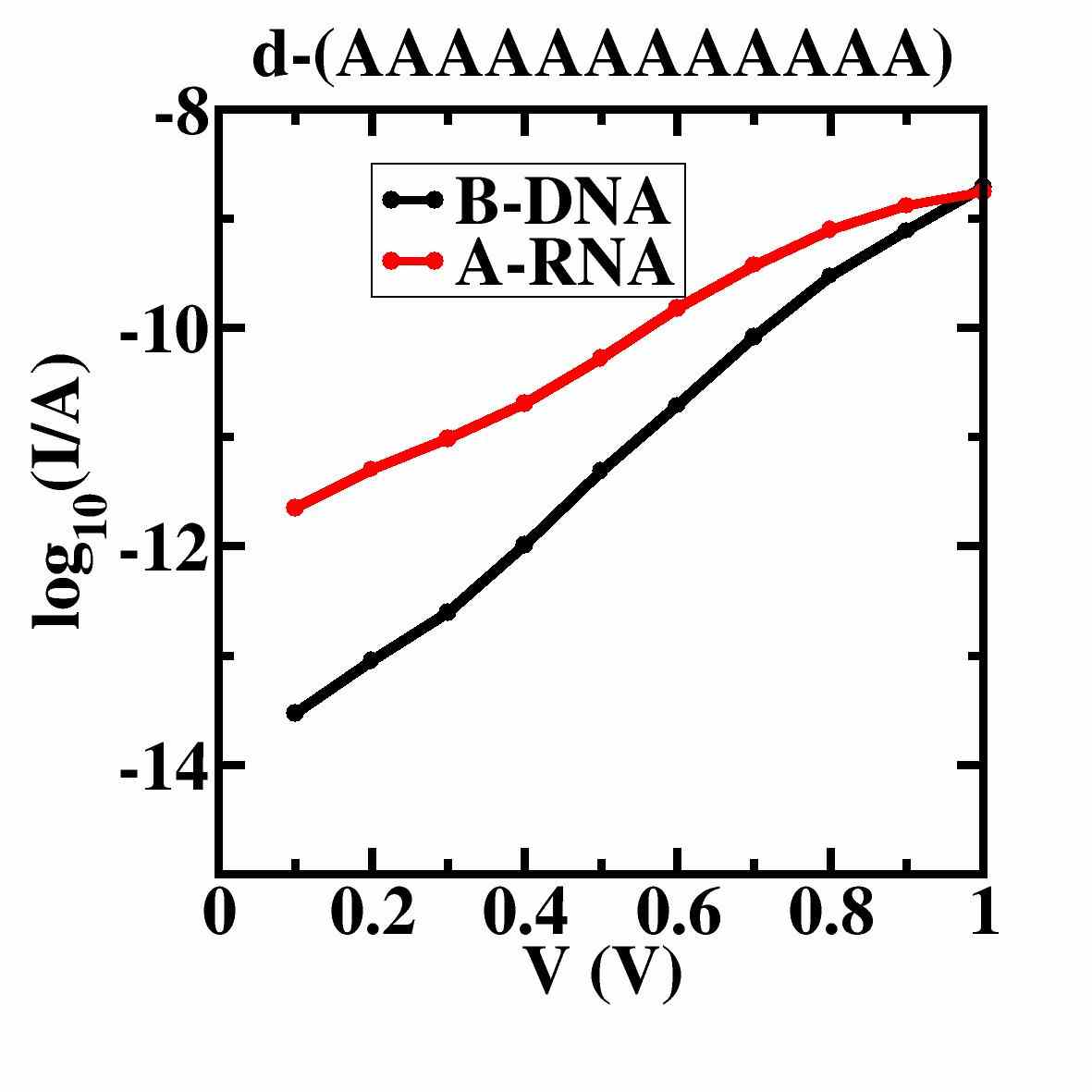}
\caption{}
    \end{subfigure}%
   \hfill
    \begin{subfigure}[t]{0.23\columnwidth}
        \centering
        \includegraphics[width=\columnwidth]{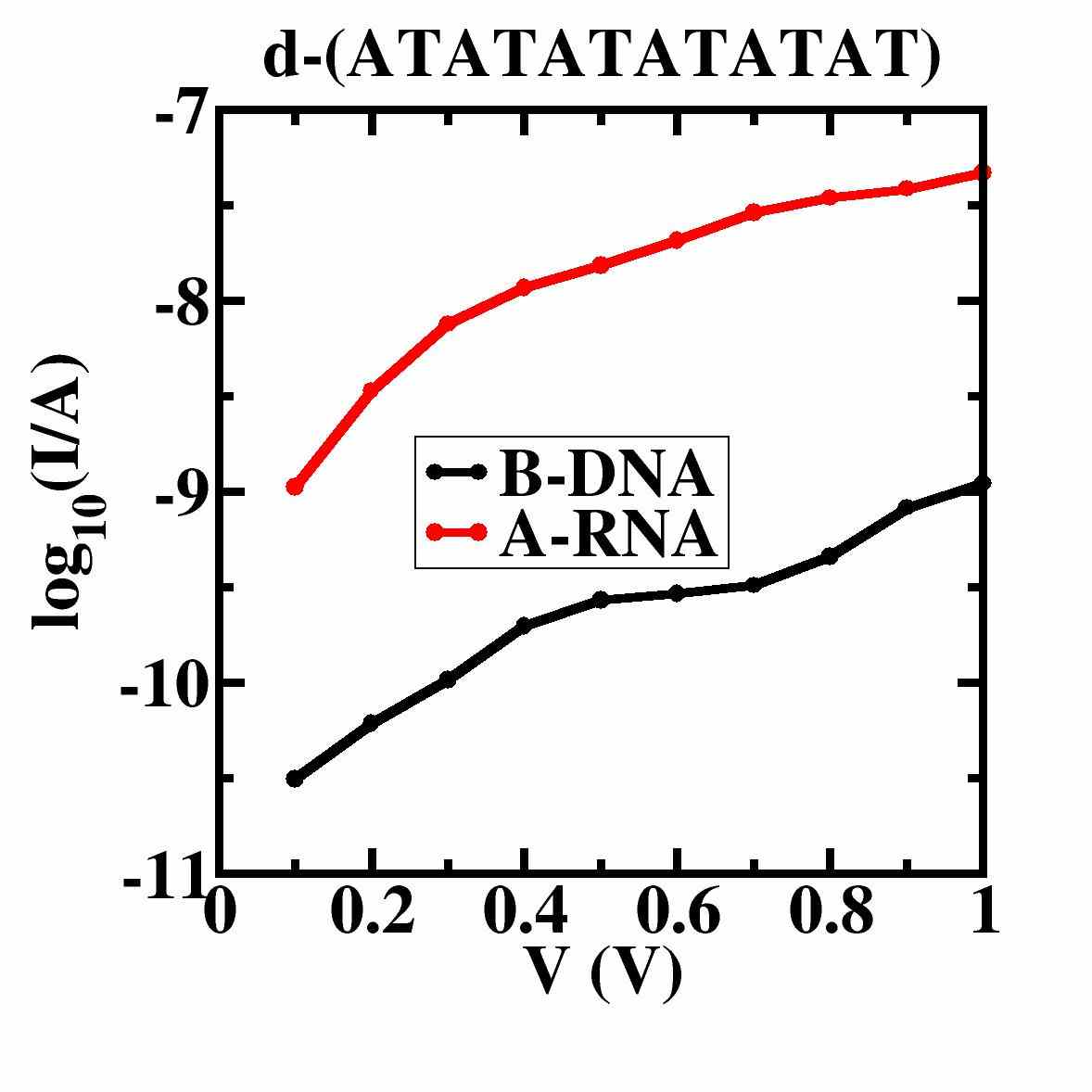}
\caption{}
\label{fig:AT_trans}
\end{subfigure}
\captionsetup{justification=raggedright, singlelinecheck=false,font=footnotesize}
	 \caption{V-I characteristics for (a) d-(CGCGCGCGCGCG) (b) d-(CCCCCCCCCCCC) (c) d-(AAAAAAAAAAAA) (d) d-(ATATATATATAT) sequences of dsDNA and dsRNA using coherent tunneling mechanism. For each sequence dsRNA conducts better than in coherent tunneling regime. }
    \label{fig:same_sequence_tunnel_current}
\end{figure}

V-I characteristics for different sequences of dsDNA and dsRNA using coherent tunneling mechanism are shown in fig. \ref{fig:same_sequence_tunnel_current}. For each sequence, dsRNA conducts better than dsDNA in coherent regime.

\clearpage

\section{ Effect of dynamic disorder in electronic couplings}

\begin{figure}[!htb]
    \centering
        \centering
            \includegraphics[width=0.5\columnwidth]{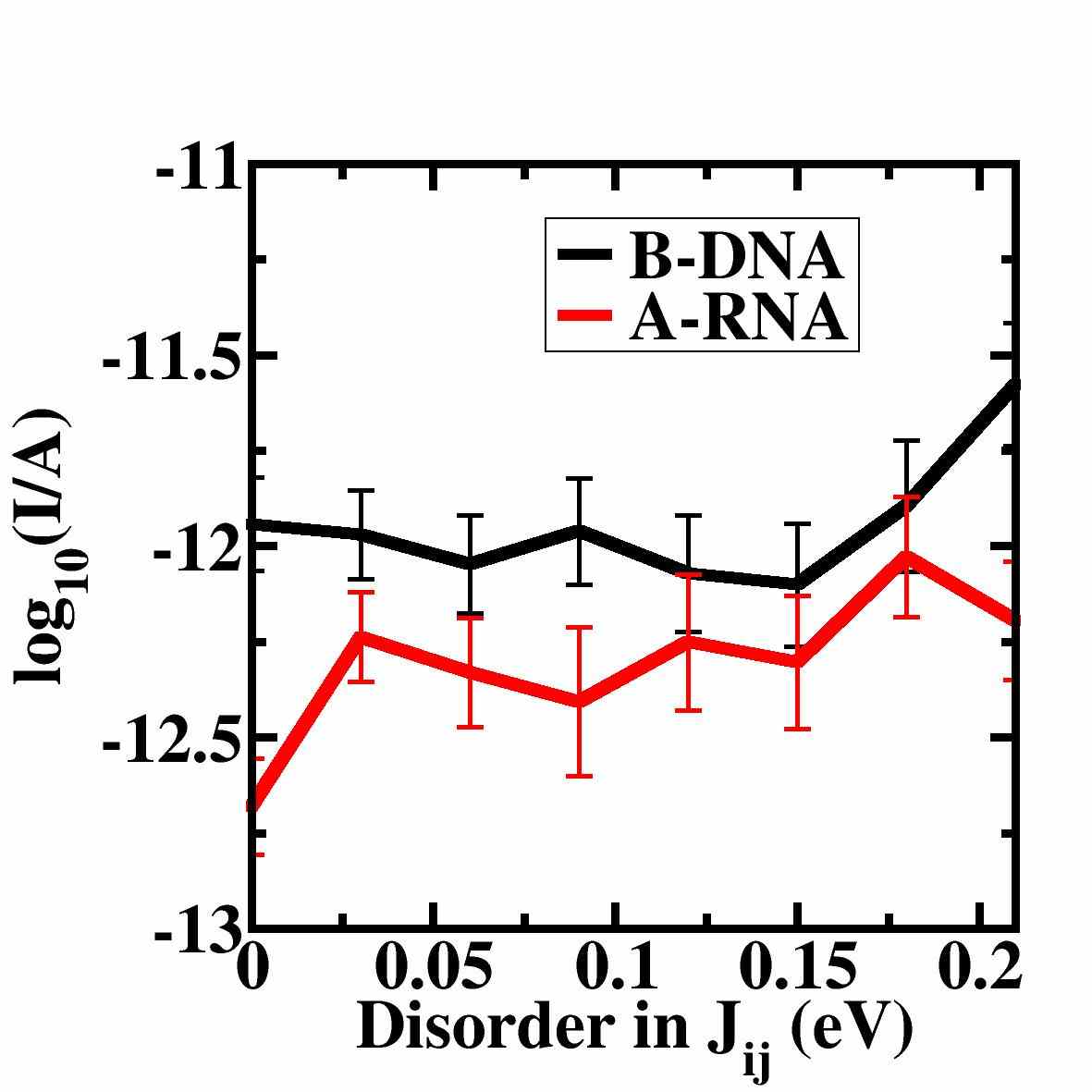}
	\caption{Variation of the current at an applied potential bias of 1 V across dsDNA and dsRNA with the amplitudes of the additional disorder incorporated to the transfer integral values. Clearly, dsDNA conducts better than dsRNA for all disorder amplitude values.}
        \label{fig:disorder}
\end{figure}

 We introduce fluctuations in the electronic couplings as uncorrelated stochastic processes in the KMC simulations.
 We sampled the additional fluctuations from a uniform distribution with amplitude, $B$, centered at zero as, $\delta J \in [-B,B] $.
Fig. \ref{fig:disorder}  shows the variation of the current at an applied potential bias of 1 V across dsDNA and dsRNA with the amplitudes of the additional disorder added to the electronic coupling values.
 Clearly, the conductance of dsDNA is higher than that of dsRNA for all values of disorders in electronic couplings considered here.

\clearpage

\section{ Experimental Results for 13 bp dsDNA and dsRNA sequence}

In order to compare the results with experiments, we also calculated the charge transport properties of 13 bp B-form dsDNA and A-form dsRNA with sequence d-(CCCGCGCGCGCCC).
The charge transport properties of this sequence have been studied experimentally in Ref. \cite{artes2015conformational}.

\begin{figure}[!htb]
    \centering
    \begin{subfigure}[t]{0.32\columnwidth}
        \centering
            \includegraphics[width=\columnwidth]{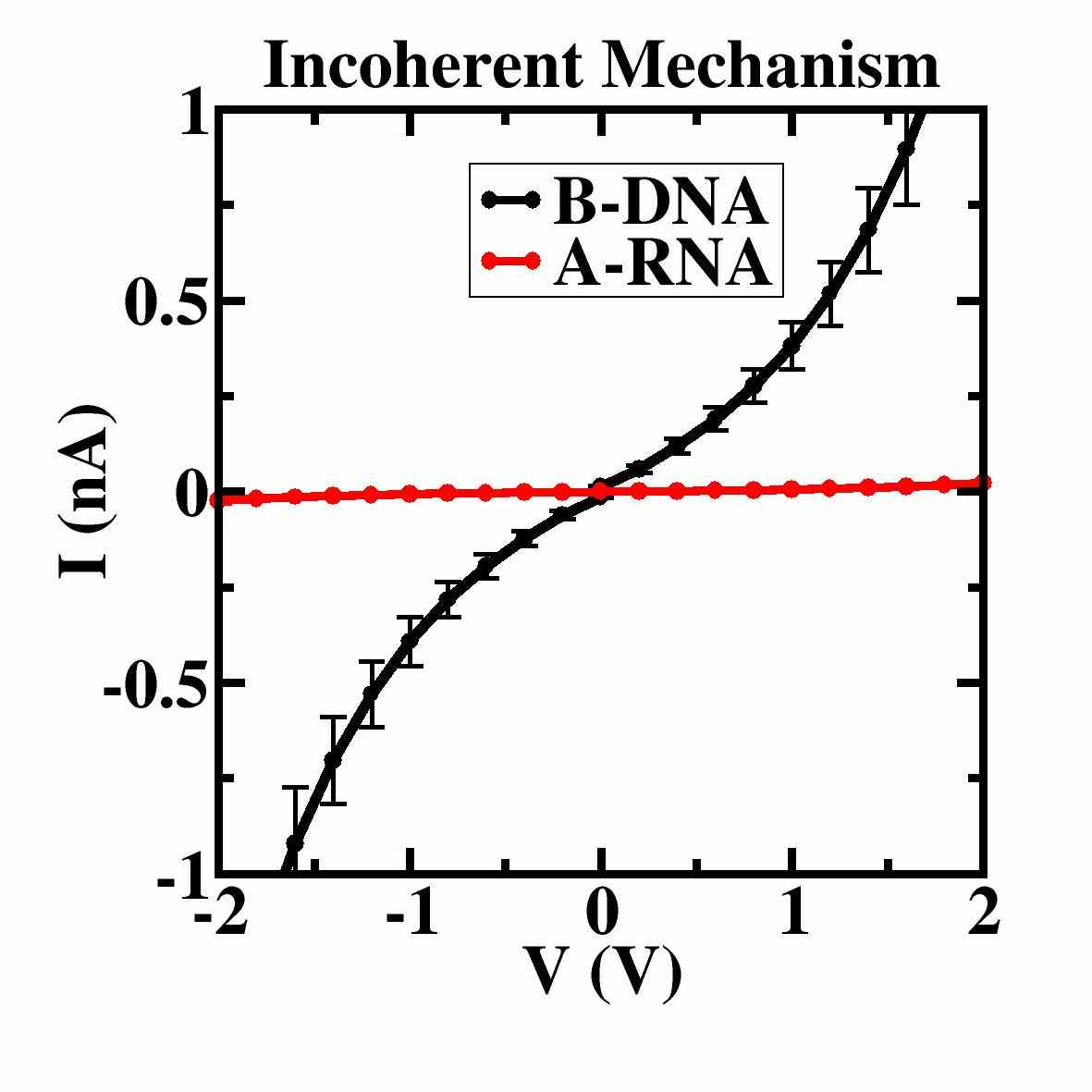}
        \caption{}
            \label{fig:9bp_hop}
    \end{subfigure}%
   \hfill
    \begin{subfigure}[t]{0.32\columnwidth}
        \centering
        \includegraphics[width=\columnwidth]{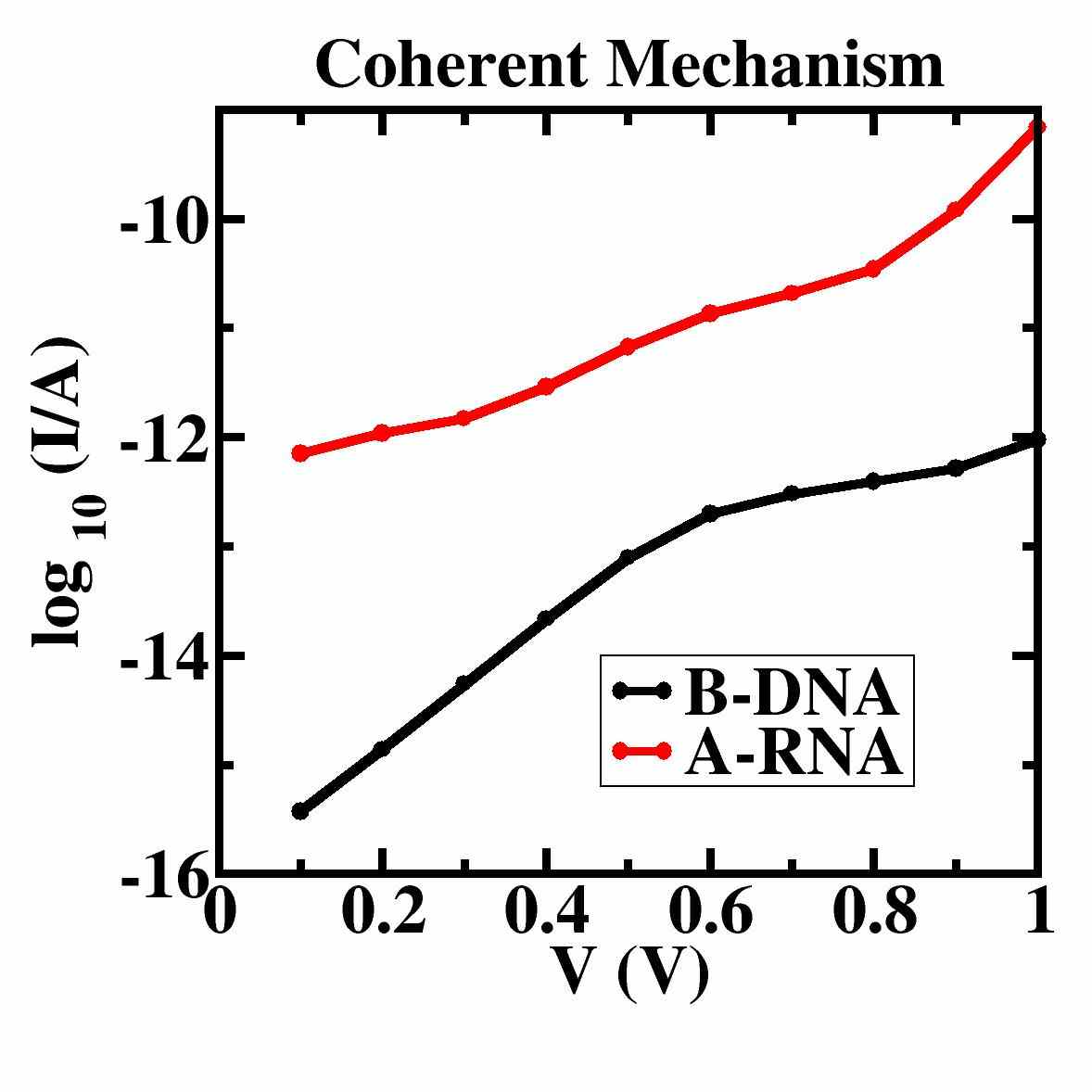}
        \caption{}
            \label{fig:9bp_lan}
    \end{subfigure}
   \hfill
    \begin{subfigure}[t]{0.32\columnwidth}
        \centering
            \includegraphics[width=\columnwidth]{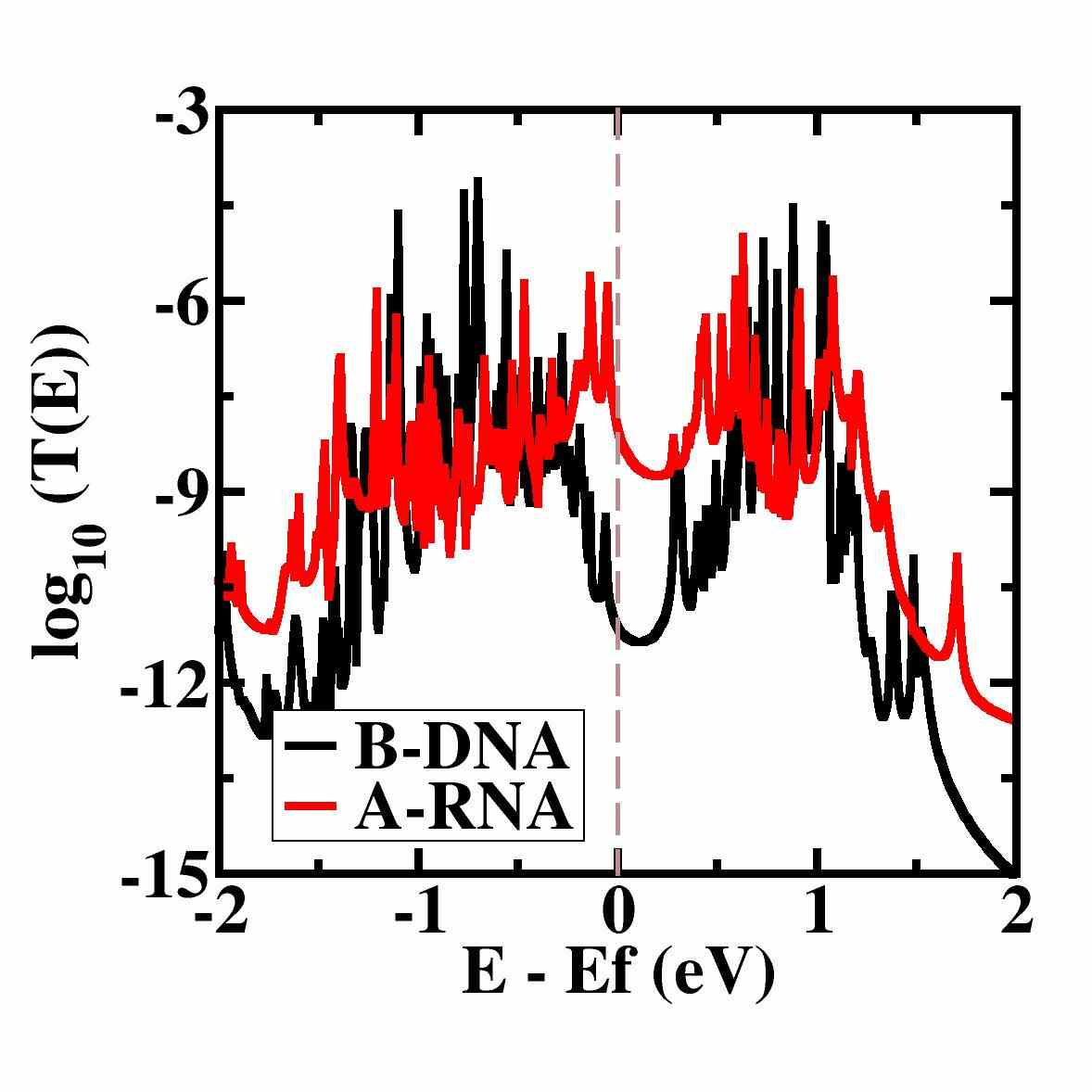}
        \caption{}
            \label{fig:9bp_tra}
    \end{subfigure}
   \hfill
	\captionsetup{justification=raggedright, singlelinecheck=false,font=footnotesize}

        \caption{(a) V-I characteristics curve for 13 bp (d-(CCCGCGCGCGCCC)) dsDNA vs dsRNA using hopping charge transport mechanism. The charge transport properties of this sequence have been studied experimentally in Ref. \cite{artes2015conformational}. dsRNA has almost two orders of magnitude less conductance than dsDNA. (b) V-I characteristics of both 13 bp (d-(CCCGCGCGCGCCC)) dsDNA and dsRNA using coherent tunneling charge transport mechanism. (c) Variation of transmission probability vs energy for 13 bp dsDNA and dsRNA. Clearly, in tunneling mechanism, dsRNA conducts better than dsDNA by an order of magnitude which is seen experimentally as well.}
    \label{fig:experiment}
\end{figure}

\clearpage

\section{ Decay constant}

\begin{figure}[!htb]
    \centering
    \begin{subfigure}[t]{0.4\columnwidth}
        \centering
        \includegraphics[width=\columnwidth]{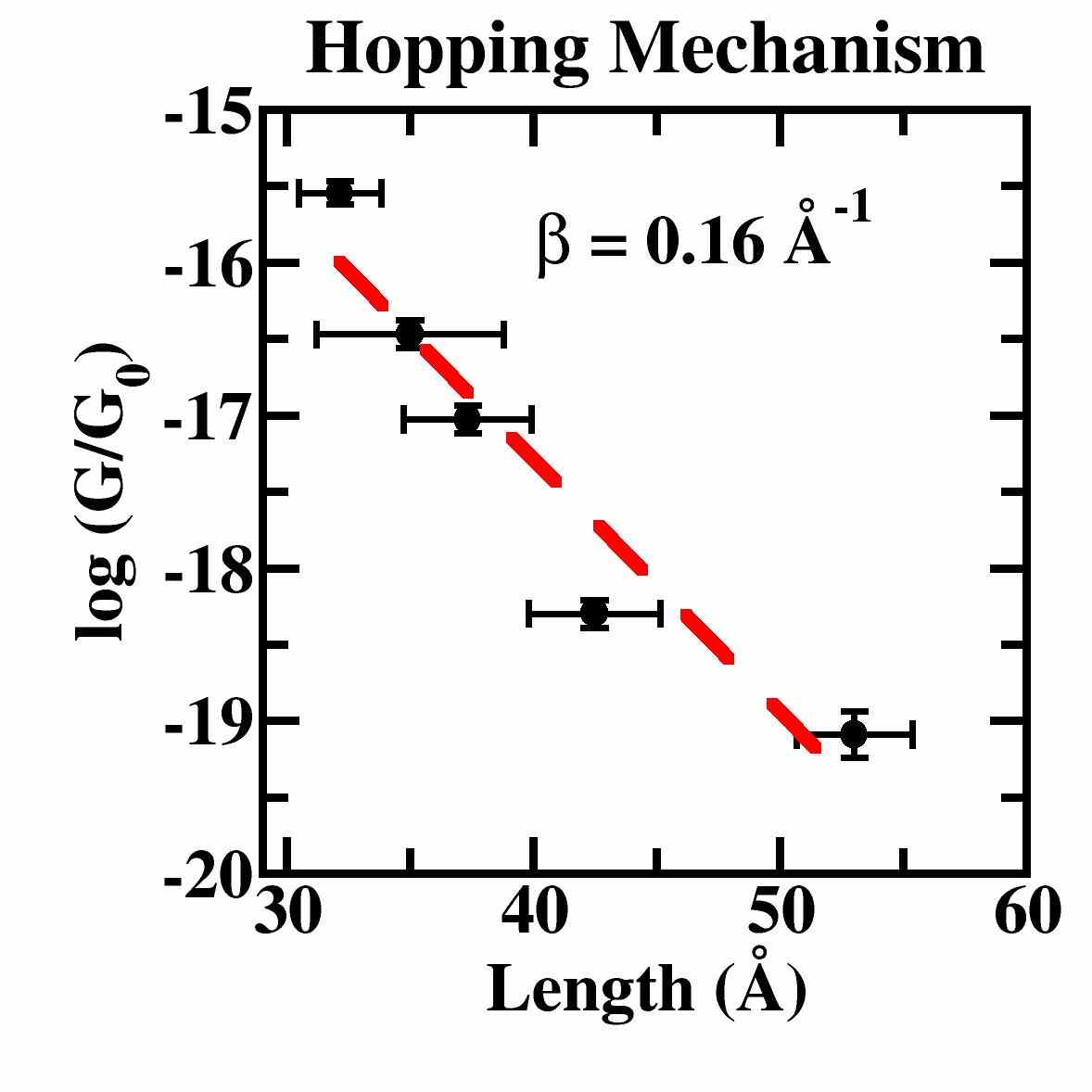}
\caption{}
	    \label{fig:decay_constant_hop}
    \end{subfigure}%
   \hfill
    \begin{subfigure}[t]{0.4\columnwidth}
        \centering
        \includegraphics[width=\columnwidth]{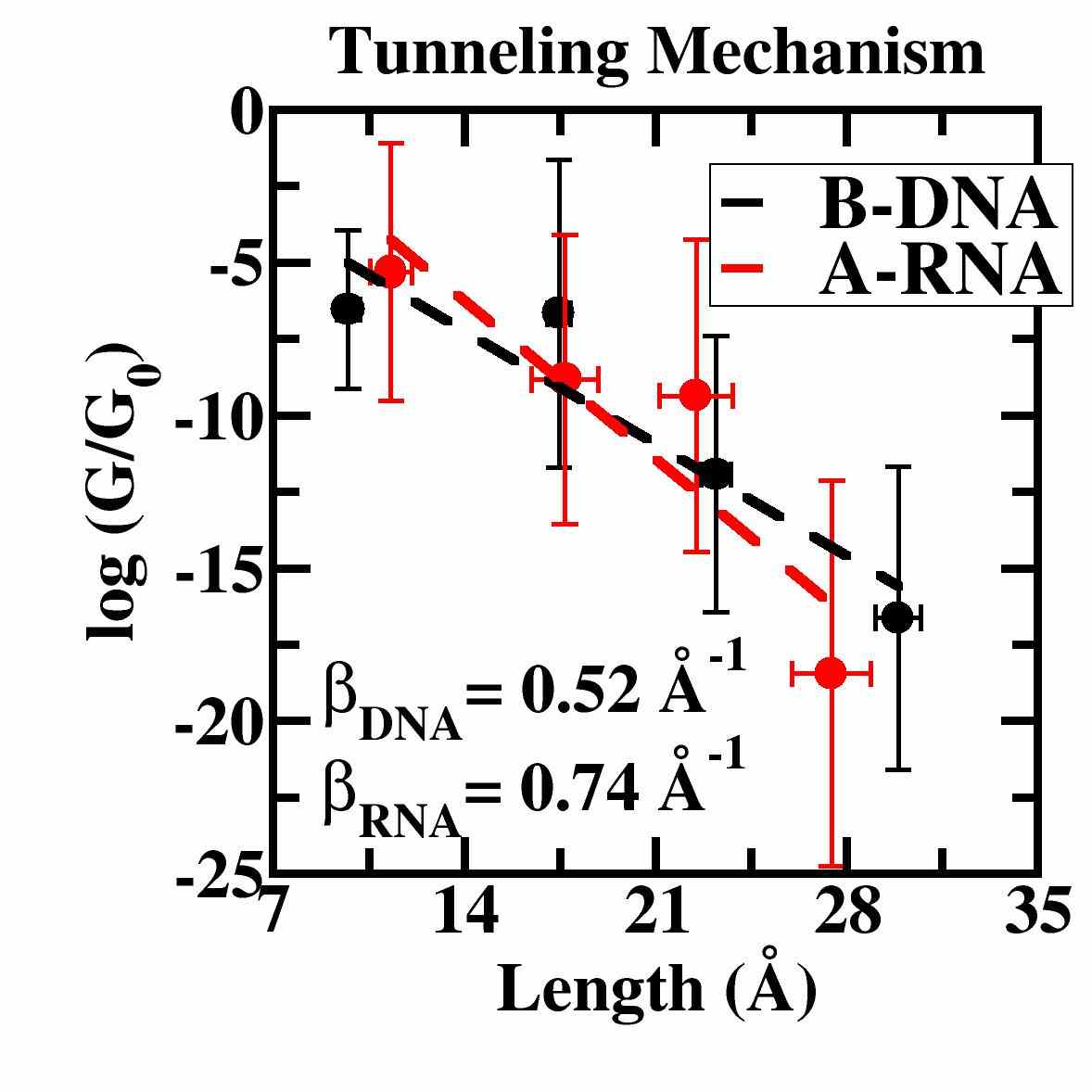}
\caption{}
\label{fig:decay_constant_tunnel}
\end{subfigure}
\captionsetup{justification=raggedright, singlelinecheck=false,font=footnotesize}
	 \caption{(a) Graph showing the variation of conductance (ln ($G/G_{0}$)) of dsRNA in hopping regime with the length of dsRNA. The decay constant of dsRNA in hopping regime is 0.16 $\AA^{-1}$.  (b)  A comparison of length dependence of dsDNA and dsRNA conductance in coherent tunneling regime. The decay constant values of dsDNA and dsRNA in tunneling regime are 0.52 $\AA^{-1}$ and $0.74 \AA^{-1}$ respectively indicating strong length dependence of current in coherent regime.}
\end{figure}

We have calculated the decay constant values for dsRNA in the hopping regime using the relation $G = G_{0} exp ^{-\beta L}$ (fig. \ref{fig:decay_constant_hop}).
 The decay constant value of 0.16 $\AA^{-1}$ implies a weak length dependence of hole current in hopping regime. 
 Whereas, the decay constant values of dsDNA and dsRNA in tunneling regime are 0.52 $\AA^{-1}$ and 0.74 $\AA^{-1}$ respectively indicating strong length dependence of current in coherent regime (fig. \ref{fig:decay_constant_tunnel}).
 Notably, the decay constant value is higher for dsRNA relative to that of dsDNA.

\clearpage

\section {Hopping model used for charge transport calculations }

\begin{figure}[t]
    \centering
    \begin{subfigure}[t]{0.4\columnwidth}
        \centering
        \includegraphics[width=\columnwidth]{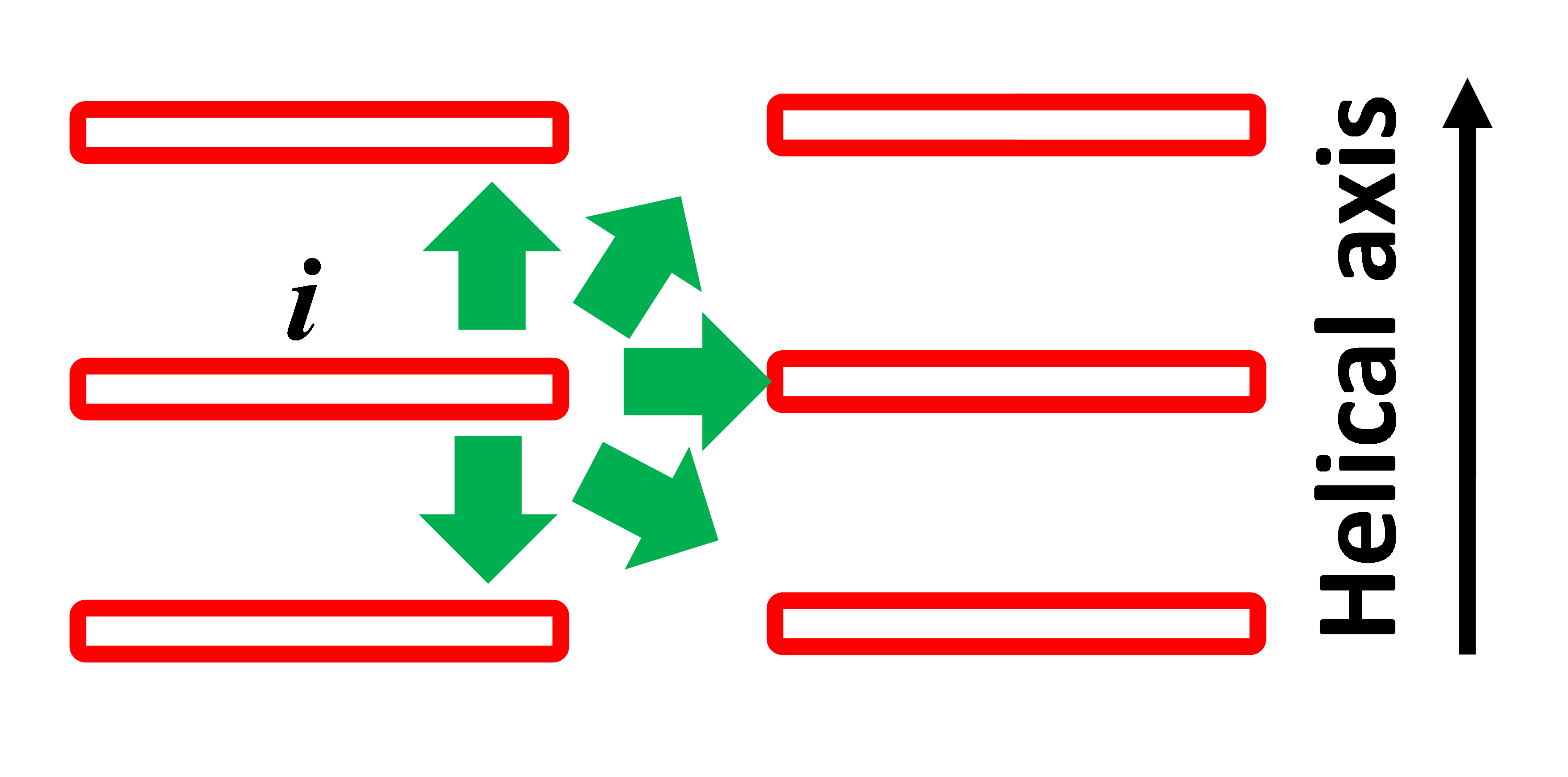}
\caption{}
    \end{subfigure}%
   \hfill
    \begin{subfigure}[t]{0.4\columnwidth}
        \centering
        \includegraphics[width=\columnwidth]{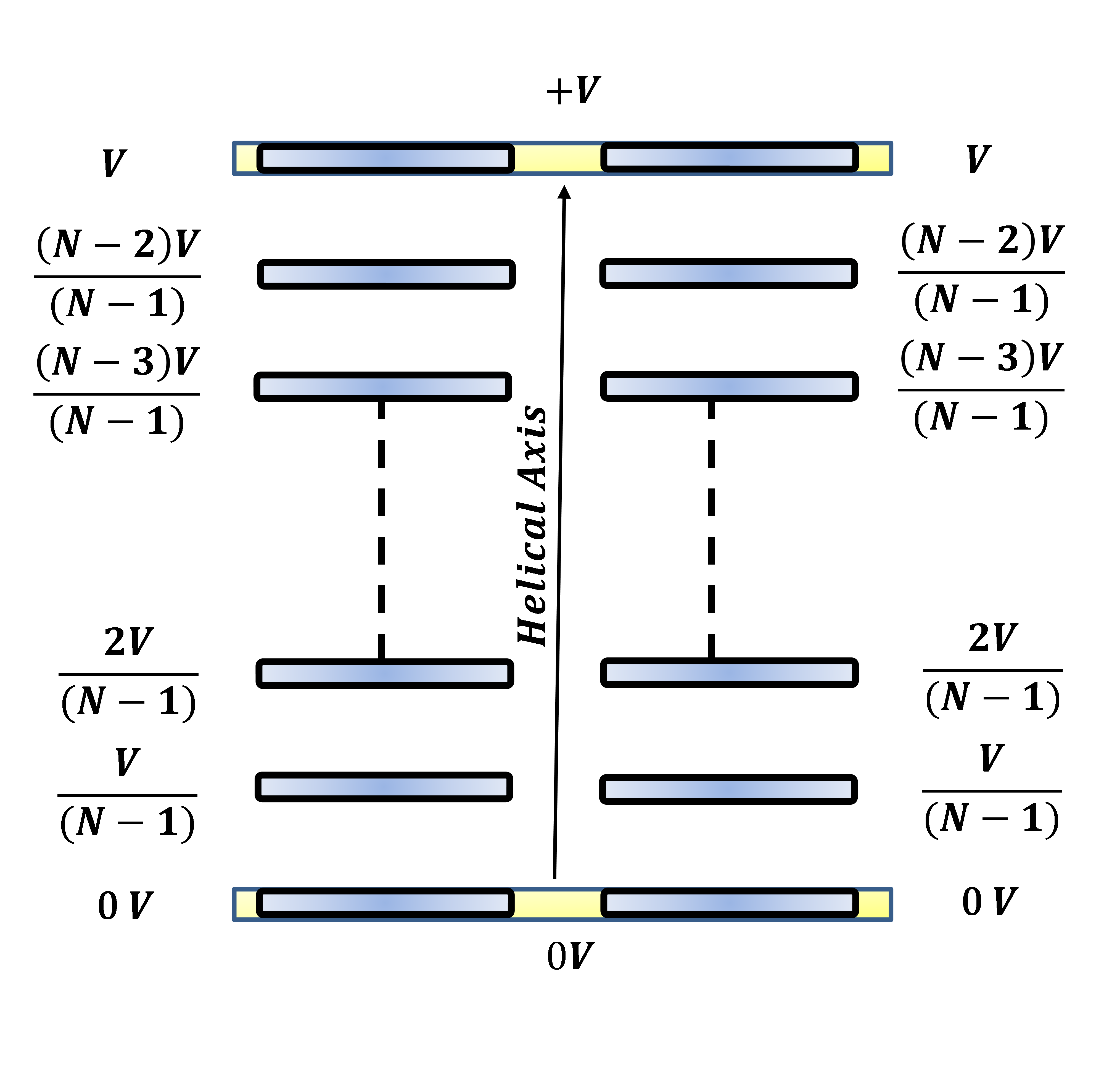}
\caption{}
\end{subfigure}
\captionsetup{justification=raggedright, singlelinecheck=false,font=footnotesize}
    \caption{(a) Schematic diagram describing the available hopping sites (i.e. the nucleobases) for a charge present at a charge hopping site $i$. A charge present at a site other than terminal bases will have 5 sites to hop to. (b) Distribution of voltage along the dsDNA or dsRNA chain during charge transport calculations. Here, we have assumed that the potential is distributed uniformly along the RNA chain.}
    \label{fig:voltage_profile}
\end{figure}

 We use nearest neighbour hopping model.
 In this model, a charge present at any hopping site other than the terminal bases has 5 available sites to hop to.
 These consists of 4 hopping sites from two adjacent base pairs and $5^{th}$ hopping site will be the complementary base of the present site.
 This is shown in fig \ref{fig:voltage_profile}.

\subsection*{Reorganization Energy and Free Energy Difference}

 $\Delta G_{ij}$ is the free energy difference between two sites defined as:
\begin{equation}
\Delta G_{ij} = \Delta G_{ij}^{int} + \Delta G_{ij}^{ext}
\end{equation}

 $\Delta G_{ij}^{ext}$ is the contribution due to the external electric field, taken as the potential difference between the two  hopping sites in our calculations.
 We take uniform distribution of potential between the base pairs, i.e. consecutive base pairs will have a potential difference of $(\frac {V}{(N-1)})$ (fig.  \ref{fig:voltage_profile}), while bases of same pairs will have zero potential difference i.e.

\begin{equation}
\Delta V = 
\begin{cases}
    (\frac {V} {N-1}), \text{for consecutive bases along the helical axis of DNA or RNA along positive voltage. } \\
    0, \text{for base pairs at the same level along the helical axis.}\\
    -(\frac {V} {N-1}), \text{for consecutive bases along the helical axis of DNA or RNA along negative voltage.}\\
\end{cases}
\end{equation}

 Whereas,

\begin{equation}
\Delta G_{ij}^{int} = U_{i}^{cC} - U_{i}^{nN} + U_{j}^{cC} - U_{j}^{nN} 
\end{equation}

 Where, $ U_{i}^{nN} (U_{i}^{cC})$ is the internal energy of neutral (charged) base in neutral (charged) state geometry.

A comparison of reorganization energy for hole and electron transfer between different base pairs dsRNA (table \ref{tab:lc}) clearly shows that the reorganization energy for electron is always higher relative to that for hole.
As the hopping rate critically depends on the reorganization energy, this difference leads to the difference in order of magnitude in hole and electron transport.

\begin{center}
\begin{table} [!htb]
\caption{Table comparing the reorganization energy for hole and electron transfer between different base pairs dsRNA }

\begin{ruledtabular}
	\begin{tabular}{||c | c | c | c||}
	From  & To & Hole (eV)  & Electron (eV)  \\
\hline

A &  A &  0.57 &   0.65 \\
A &  C &  0.46 &   0.84 \\
A &  G &  0.64 &   0.73 \\
A &  U &  0.55 &   0.73 \\
C &  A &  0.54 &   0.93 \\
C &  C &  0.43 &   1.12 \\
C &  G &  0.61 &   1.01 \\
C &  U &  0.51 &   1.01 \\
G &  A &  0.61 &   1.20 \\
G &  C &  0.50 &   1.39 \\
G &  G &  0.68 &   1.28 \\
G &  U &  0.58 &   1.28 \\
U &  A &  0.60 &   0.98 \\
U &  C &  0.49 &   1.17 \\
U &  G &  0.67 &   1.06 \\
U &  U &  0.57 &   1.06 \\

\end{tabular}
\end{ruledtabular}
\label{tab:lc}
\end{table}
\end{center}

\clearpage

\section*{Effect of external Reorganization Energy on hopping charge transport of dsDNA and dsRNA}
\begin{figure}[!htb]
    \centering
        \includegraphics[width=0.7\columnwidth]{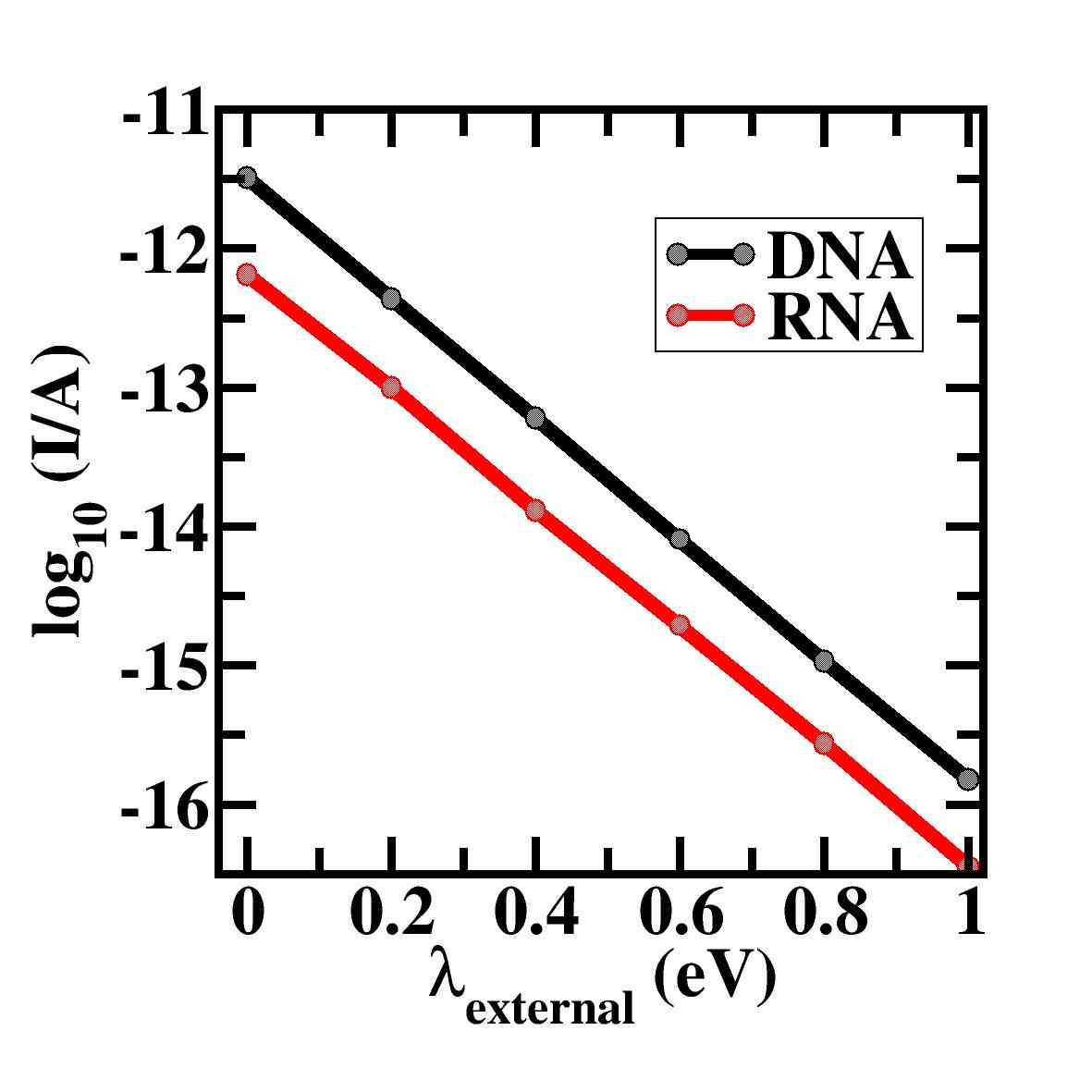}
\captionsetup{justification=raggedright, singlelinecheck=false,font=footnotesize}
	\caption{Variation of current at an applied potential bias of 1 V with outer-sphere reorganization energy for 12 bp dsDNA (d-(CGCGAATTCGCG)) and dsRNA (d-(CGCGAAUUCGCG)). Clearly, dsDNA conducts better than dsRNA for all values of outer-sphere reorganization energies considered here.}
    \label{fig:outer}
\end{figure}

 The external reorganization energy, $\lambda_{ij}$, has two parts : inner sphere reorganization energy and outer sphere reorganization energy.
 The outer sphere reorganization energy is varied from 0 eV to 1 eV and the effect on the current at an applied potential bias of 1 V through dsDNA and dsRNA has been plotted in fig. \ref{fig:outer}. 
 Clearly, dsDNA conducts better relative to dsRNA for all outersphere reorganization energy values. 
 Hence, outersphere reorganization energy is taken as 0 in all the calculations.

\clearpage

\subsection*{Ionization Energies and Electron Affinities of DNA base pairs}
 Table \ref{tab:ie_ea} lists the ionization energies and electron affinities of DNA and RNA nucleobases.

\begin{table}[!htb]
\caption{Table comparing the ionization energies and electron affinities of DNA and RNA nucleobases }

\begin{ruledtabular}
        \begin{tabular}{||c | c | c||}
        Base & Ionization Energy (eV)  & Electron Affinity (eV)  \\
\hline

 A & 6.15 &  -0.93  \\
 C & 6.51 &  -1.32  \\
 G & 5.74 &  -0.61  \\
 T & 6.80 &  -1.42  \\
 U & 6.51 &  -1.33  \\

\end{tabular}
\end{ruledtabular}
\label{tab:ie_ea}
\end{table}

\bibliography{references}
